\documentclass[epsf,useAMS,usenatbib, usegraphicx]{mn2e}
\usepackage{graphicx}
\usepackage{epsfig}
\usepackage{amsmath}
\usepackage{amssymb}
\usepackage{rotating}
\usepackage{color}
\usepackage{pifont}
\usepackage{longtable}
\usepackage{pdflscape,lscape}
\newcommand{\ion}[2]{{#1}\,{\sc #2}}
\newcommand{\teff}{$T_{\rm eff}$}
\newcommand{\logg}{$\log g$}

\newcommand{\kms}{km\,s$^{-1}$}
\makeatletter
\def\@xfootnote[#1]{%
  \protected@xdef\@thefnmark{#1}%
  \@footnotemark\@footnotetext}
\makeatother

\title[Spectroscopic Survey of $\gamma$\,Doradus Stars]{Spectroscopic Survey of $\gamma$\,Doradus Stars \\
I. Comprehensive atmospheric parameters and abundance analysis of $\gamma$\,Doradus stars}

\author[F. Kahraman Ali\c{c}avu\c{s} et al.]{F. Kahraman Ali\c{c}avu\c{s}$^{1,2}$\thanks{E-mail: filizkahraman01@gmail.com}, E. Niemczura$^{2}$, 
P. De Cat$^{3}$, E. Soydugan$^{1}$,\and Z. Ko\l{}aczkowski$^{2}$, J. Ostrowski$^{2}$, J. H. Telting$^{4}$, K. Uytterhoeven$^{5,6}$, 
E. Poretti$^{7}$,\and M. Rainer$^{7}$, J. C. Su\'{a}rez$^{8}$, L. Mantegazza$^{7}$, P. Kilmartin$^{9}$ and K. R. Pollard$^{9}$
\\
$^{1}$Faculty of Sciences and Arts, Physics Department,Canakkale Onsekiz Mart University, 17100, Canakkale, Turkey\\
$^{2}$Instytut Astronomiczny, Uniwersytet Wroc\l{}awski, Kopernika 11, 51-622 Wroc\l{}aw, Poland\\
$^{3}$Royal observatory of Belgium, Ringlaan 3, B-1180 Brussel, Belgium \\
$^{4}$Nordic Optical Telescope, Rambla Jos\'e Ana Fern\'andez P\'erez 7, 38711 Bre\~na Baja, Spain\\
$^{5}$Instituto de Astrofisica de Canarias, E-38205  La Laguna, Tenerife, Spain\\
$^{6}$Universidad de La Laguna, Departamento de Astrofisica, E-38206  La Laguna, Tenerife, Spain\\
$^{7}$INAF-Osservatorio Astronomico di Brera, Via Bianchi 46, I-23807 Merate, Italy\\
$^{8}$Department F\'{\i}sica Te\'{o}rica y del Cosmos. Universidad de Granada. Campus de Fuentenueva, 18071. Granada. Spain\\
$^{9}$Department of Physics and Astronomy, University of Canterbury, Private Bag 4800, Christchurch, New Zealand }

\begin{document}

\date{Accepted ... Received ...; in original form ...}

\pagerange{\pageref{firstpage}--\pageref{lastpage}} \pubyear{2016}

\maketitle

\label{firstpage}

\begin{abstract}
We present a spectroscopic survey of known and candidate $\gamma$\,Doradus stars. 
The high-resolution, high signal-to-noise spectra of 52 objects were collected by five different spectrographs.
The spectral classification, atmospheric parameters (\teff, $\log g$, $\xi$), $v\sin i$ and chemical composition of the stars were derived.
The stellar spectral and luminosity classes were found between G0-A7 and IV-V, respectively.
The initial values for \teff\ and \logg\ were determined from the photometric indices and spectral energy distribution.
Those parameters were improved by the analysis of hydrogen lines.
The final values of \teff, \logg\ and $\xi$ were derived from the iron lines analysis. 
The \teff\ values were found between 6000\,K and 7900\,K, while \logg\,values range from 3.8 to 4.5\,dex. 
Chemical abundances and $v\sin i$ values were derived by the spectrum synthesis method. 
The $v\sin i$ values were found between 5 and 240\,km\,s$^{-1}$.
The chemical abundance pattern of $\gamma$\,Doradus stars were compared with the pattern of non-pulsating stars. 
It turned out that there is no significant difference in abundance patterns between these two groups. 
Additionally, the relations between the atmospheric parameters and the pulsation quantities were checked. 
A strong correlation between the $v\sin i$ and the pulsation periods of $\gamma$\,Doradus variables was obtained. 
The accurate positions of the analysed stars in the H-R diagram have been shown.
Most of our objects are located inside or close to the blue edge of the theoretical instability strip of $\gamma$\,Doradus.

\end{abstract}


\section{Introduction}

The class of $\gamma$\,Doradus ($\gamma$\,Dor) variables was defined by \citet{1994MNRAS.270..905B} 
after discovery of the variability of the prototype of these pulsators \citep{1992Obs...112...53C, 1993MNRAS.263..781K}.
The $\gamma$\,Dor variables exhibit pulsations in the non-radial, high-order (\textit{n}), low-degree (\textit{l}) gravity modes 
with amplitudes at the level of 0.1\,mag (V) and pulsation periods between 0.3 and 3 days \citep{1999PASP..111..840K}.
The pulsations of $\gamma$\,Dor stars are driven by the mechanism of convective flux blocking \citep{2000ApJ...542L..57G, 2005A&A...435..927D}.
In the Hertzsprung-Russell (H-R) diagram, the theoretical instability strip of $\gamma$\,Dor variables is located partially inside the instability strip 
of $\delta$\,Scuti ($\delta$\,Sct) stars. In this small overlapping part, the existence of stars pulsating simultaneously in both $\delta$\,Sct 
and $\gamma$\,Dor modes was predicted \citep{2004A&A...414L..17D}. These stars are called $\gamma$\,Dor/$\delta$\,Sct or A-F type hybrids.
The $\gamma$\,Dor variables are A7-F5 dwarfs and/or sub-dwarfs \citep{1999PASP..111..840K}. 
This means that in the H-R diagram they are situated inside the region of the transition from a convective envelope to a convective core. 
In order to reveal properties of the pulsation mechanism in F type stars and to decide on the correct location of 
the theoretical instability strip of $\gamma$\,Dor stars in the H-R diagram the interaction between convection and pulsation 
has to be taken into account \citep{2015MNRAS.447.3264S}. 
Moreover, the investigation of $\gamma$\,Dor variables allows us to examine important subjects of the internal structure and evolution 
of intermediate mass stars \citep{2008MNRAS.386.1487M}. In particular, the frequency spacing detected 
in the photometric time series has allowed the study of the internal structure and surface-to-core rotation
\citep [e.g.][]{2014MNRAS.444..102K, 2015ApJS..218...27V}.

In-depth studies of the pulsating A-F type stars have now become possible due to the space observations.
In particular, the high-precision light curves of the {\sl Kepler} mission enabled investigation of many new A-F type variables \citep{2010Sci...327..977B}.
Before the space observations, approximately 100 $\gamma$\,Dor stars were known \citep{2011AJ....142...39H}.
The analysis of the {\sl Kepler} data revealed many new candidate $\gamma$\,Dor, $\delta$\,Sct and A-F type hybrid stars \citep{2011A&A...534A.125U, 2010ApJ...713L.192G}.
The investigation of {\sl Kepler} observations and ground-based photometric data allow us to determine 
pulsation characteristics, ranges of fundamental parameters, and position of these variables in the H-R diagram.

However, many new questions about the properties of the $\gamma$\,Dor stars, $\delta$\,Sct stars and their hybrids arose.
The first question concerns the exact location of the instability domains of these variables in the H-R diagram.
According to the existing studies, there seems to be no clear distinction in the edges of the observational instability strip of 
$\gamma$\,Dor and $\delta$\,Sct stars. 
Moreover, it was shown that candidate hybrids of $\gamma$\,Dor and $\delta$\,Sct stars were detected everywhere inside 
the theoretical instability strips of both types of pulsating stars \citep[e.g.][]{2014MNRAS.444..102K, 2015MNRAS.450.2764N}.
Another question relates to the chemical structure of the hybrid stars.
Some Am hybrid stars were discovered, and these results showed that a relation between the Am phenomenon and hybridity could exist \citep{2011ApJ...743..153H}.
Solving these problems requires investigation of whether chemical and physical differences between hybrids, $\gamma$\,Dor, and $\delta$\,Sct variables exist.
Therefore, it is necessary to obtain the accurate physical and chemical characteristics of all classes of A-F type variables.
Hence, reliable spectroscopic and multi-colour photometric studies are essential.

So far, several photometric and spectroscopic studies of $\gamma$\,Dor stars have been carried out \citep[e.g.][]{2007AJ....133.1421H, 2004A&A...417..189M}.
One of the most detailed spectroscopic investigations of $\gamma$\,Dor stars was presented by \citet{2008A&A...478..487B}.
They derived fundamental atmospheric parameters and chemical composition of \textit{bona-fide} and candidate $\gamma$\,Dor stars to 
search for links between $\gamma$\,Dor, Am and $\lambda$ Bo\"{o}tis stars, but no relations were found.
Additionally, detailed spectroscopic analyses of $\gamma$\,Dor stars detected in satellite 
fields have been carried out \citep[e.g.][]{2012MNRAS.422.2960T, 2013MNRAS.431.3685T, 2015MNRAS.450.2764N, 2015ApJS..218...27V}.
In these studies, the fundamental atmospheric parameters and chemical abundances of these variables were derived.

The aim of this study is to obtain the atmospheric parameters and chemical abundances of some \textit{bona-fide} 
and candidate $\gamma$\,Dor stars detected from the ground-based observations.
Therefore, we gathered high-resolution and high signal-to-noise (S/N) spectra for 69 $\gamma$\,Dor stars 
using five different spectrographs from around the world.
This sample contains a mixture of single stars, single-lined binaries (SB1) and double-lined spectroscopic binaries (SB2).
The analysis of SB2 $\gamma$\,Dor stars will be presented in a separate paper (Kahraman Ali\c{c}avu\c{s} et al., in preparation).
In this study, a detailed spectroscopic analysis of 52 single and SB1 $\gamma$\,Dor stars is performed.
The high-resolution observation, data reduction and calibration details are given in Sect.\,2. 
Spectral classification process is described in Sect.\,3. Determination of the atmospheric parameters from photometric systems 
and spectral energy distribution are presented in Sect.\,4. In Sect.\,5, we introduce the atmospheric parameters determination 
from the analysis of hydrogen and iron lines, the detailed chemical abundance analysis, and discussion of the obtained parameters. 
The summary of the results and an outlook for future studies are given in Sect.\,6.

\begin{table*}
\centering
  \caption{Information about the spectroscopic observations.}
  \begin{tabular}{@{}lcccccc@{}}
  \hline
Instrument  & Number of                        & Years of       & Resolving & Spectral      & S/N \\
            & single \& SB1/SB2 stars  & observations   & power     & range [{\AA}] & range \\
 \hline
 FEROS    & 0\,/\,8 & 2008      & 48000  & 3500-9200 &170\,-\,340 \\
 FIES     & 29\,/\,2& 2007-2010 & 67000  & 3700-7300 &100\,-\,330\\
 HARPS    & 11\,/\,4& 2009-2011 & 80000  & 3780-6910 &130\,-\,360\\
 HERCULES & 11\,/\,3& 2007-2010 & 70000  & 4000-8800 &110\,-\,300\\
 HERMES   & 1\,/\,0 & 2010      & 85000  & 3770-9000 &150\\
\hline
\end{tabular}
\end{table*}

\section[]{Observations}

The observations of our targets were carried out with five high-resolution spectrographs.
Numbers of the observed single \& SB1 and SB2 stars, observation years, spectral resolutions of instruments, 
wavelength range and signal-to-noise (S/N) ranges are given in Table\,1. 
For each spectrograph, the listed S/N range gives the maximum and minimum value of S/N at 5500\,{\AA}.
The following instruments were used in the survey:
\begin{description}

\item[--] \textit{FEROS} (Fibre-fed Extended Range Optical Spectrograph), an \'{e}chelle spectrograph attached to the 2.2-m telescope of 
the European Southern Observatory (ESO, La Silla, Chile) 
\citep{2012MNRAS.420.2727E};

\item[--] \textit{FIES} (Fibre-fed \'{E}chelle Spectrograph), a cross-dispersed high-resolution \'{e}chelle spectrograph attached to the 
2.56-m Nordic Optical Telescope of the Roque de los Muchachos Observatory (ORM, La Palma, Spain) 
\citep{2014AN....335...41T}; 

\item[--] \textit{HARPS} (High Accuracy Radial Velocity Planet Searcher), an \'{e}chelle spectrograph attached to the 3.6-m telescope 
of the European Southern Observatory (ESO, La Silla, Chile) \citep{2003Msngr.114...20M};

\item[--] \textit{HERCULES} (High Efficiency and Resolution Canterbury University Large \'{E}chelle Spectrograph), a fibre-fed 
\'{e}chelle spectrograph attached to the 1-m McLellan telescope of the Mt. John University Observatory (MJUO, Mount John, New Zealand)
\citep{2003ASPC..289...11H};

\item[--] \textit{HERMES} (High Efficiency and Resolution Mercator \'{E}chelle Spectrograph), a high-resolution fibre-fed 
\'{e}chelle spectrograph attached to the 1.2-m Mercator telescope at the Roque de los Muchachos Observatory (ORM, La Palma, Spain) 
\citep{2011A&A...526A..69R}.

\end{description}

The collected spectra have been reduced and calibrated using the dedicated reduction pipelines of the instruments.
The usual reduction steps for \'{e}chelle spectra were applied, i.e.: bias subtraction, flat-field correction, 
removal of scattered light, order extraction, wavelength calibration, and merging of the orders.
For the HERCULES data, an additional procedure had to be used to merge the \'{e}chelle orders. 
In this procedure the overlapping parts of the orders were averaged using the signal-to-noise of the given order as the weight.
The normalisation of all spectra was performed manually by using the \textit{continuum} task of the NOAO/IRAF package\footnote{http://iraf.noao.edu/}.

Some of the studied stars were observed by more than one instrument.
In this case, only the spectra of the instrument with the highest resolution were analysed.
For some of the stars we collected more than one spectrum from the same instrument.
For these stars all the spectra were combined and the average spectrum was investigated.

We collected the spectroscopic observations of both single-lined (single stars and SB1 binaries) and double-lined (SB2 binaries) stars.
Some of these spectroscopic binaries had already been known in the literature as SB2 objects.
In our sample, four new SB2 systems were detected: HD\,85693, HD\,155854, HD\,166114 and HD\,197187.
The number of spectra we have so far for these targets is insufficient to determine their orbits.
In this paper, we present the spectroscopic analysis of single and single lined spectroscopic binaries (SB1) only.
An overview of the analysed objects is given in Table\,2.

\section{Spectral classification}

A spectral classification gives crucial information about chemical peculiarity and initial atmospheric parameters of a star.
Determination of the spectral type and the luminosity class relies on a comparison of the spectra of the studied stars with 
those of well-known standards, taking into account important hydrogen and metal lines.

As the $\gamma$\,Dor stars are late-A to mid-F type stars, we used only the spectra of standard A and F type stars from 
\citet{2003AJ....126.2048G} in the classification process. 
For each star, the spectral type determination was carried out three times, each time based on a different set of lines:
\\
(1) \textit{Hydrogen lines}: H$\gamma$ and H$\delta$ lines,\\
(2) \textit{Metal lines}: \ion{Fe}{i}, \ion{Ca}{i} and \ion{Mg}{i} and their ratios with the Balmer lines,\\
(3) \textit{\ion{Ca}{ii}\,K line} (stars earlier than F3) or \textit{G-band} (for late F type stars).\\
In the case of a non-chemically peculiar star, all three methods should give the same result.
However, for chemically peculiar Am or $\lambda$\,Bo\"{o}tis stars, 
different spectral types are obtained from different sets of lines \citep{2009ssc..book.....G}.

To obtain luminosity classes, we used blended lines of ionised iron and titanium near 4500\,{\AA} \citep{2009ssc..book.....G}.
In the case of A and early F type stars, Balmer lines are good indicators of the luminosity class while the G-band can be used for late F type stars. 
The luminosity classes were determined using all these indicators.

The resulting spectral types and luminosity classes are given in Table\,2.
They range between A7 and G0, and between IV and V, respectively.
In the classification process, we discovered two mild Am stars, showing a difference of less than 
five spectral subtypes in the results based on the metal lines and 
the \ion{Ca}{ii}\,K line: HD\,33204 (kA7\,hA7\,mF2\,V) and HD\,46304 (kA7\,hA8\,mF0\,V). 
These mild Am stars are denoted as 'Am\,:' in Table\,2.
HD\,33204 was already classified as an Am star (kA5\,hA7\,mF2) by \citet{1989ApJS...70..623G}. 
We also found metal poor stars, exhibiting weak metal lines. 
These metal poor objects are indicated by 'm\,-*'.
This notation represents the metallicity spectral class where '*' is a number: e.g. F2\,m\,-\,2 means that metallicity spectral 
class of this star is F0 \citep{2009ssc..book.....G}. 
 
\section{Stellar parameters from photometry and SED}

Before the analysis of the high-resolution spectra, we derived initial values for atmospheric parameters of the stars 
using both different photometric indices (Sect.\,4.1) and the SED method (Sect.\,4.2). 
However, photometric colours and SEDs are very sensitive to the interstellar reddening (${\it E(B-V)}$). 
Therefore, values of the interstellar reddening were first calculated using two different approaches.

In the first method, we used the interstellar extinction map code written by Dr. Shulyak (private information)  
based on the Galactic extinction maps published in \citet{2005AJ....130..659A}.
The ${\it E(B-V)}$ values from the extinction maps were calculated using the \textit{Hipparcos} parallaxes \citep{2007A&A...474..653V} and 
stellar galactic coordinates from the SIMBAD data base \citep{2000A&AS..143....9W}\footnote{http://simbad.u-strasbg.fr/simbad/}. 
Because of the lack of parallaxes for cluster members HD\,22702 (Melotte 22 3308) and HD\,169577 (NGC\,6633 15),  
their distances were assumed as cluster distances, being 130 and 385\,pc \citep{2005A&A...438.1163K}, respectively.

In the second method, we derived ${\it E(B-V)}$ values from interstellar sodium lines.
This approach is based on the relation between the equivalent width of the Na\,D$_{2}$ line (5889.95 {\AA}) 
and the ${\it E(B-V)}$ value \citep{1997A&A...318..269M}.

\begin{figure}
\includegraphics[width=8.5cm, angle=0]{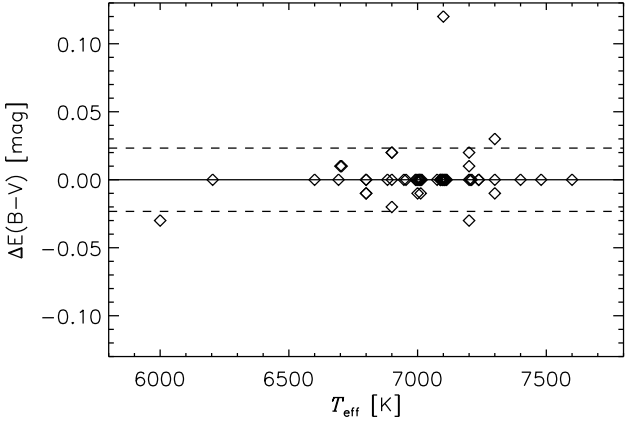}
\caption{The differences between the ${\it E(B-V)}$ values derived from interstellar maps and the sodium Na\,D$_{2}$ lines. 
The dashed lines show 1-$\sigma$ level.}
\label{figure1}
\end{figure}

The resulting ${\it E(B-V)}$ values obtained with both methods are listed in Table\,3 and compared with each other in Fig.\,\ref{figure1}. 
Uncertainties of ${\it E(B-V)}$ values were adopted as equal 0.02\,mag on the basis of the standard deviation resulting from the 
comparison of the values from two methods (1-$\sigma$, see Fig.\,\ref{figure1}).
It can be seen that the results are consistent with each other, except for HD\,169577. 
For this star, the difference between both values is about 0.1\,mag. 
Note that for this star we used the NGC\,6633 cluster distance in our calculation (first method). 
For the determination of stellar parameters, the average ${\it E(B-V)}$ values of both methods were used.

\begin{figure*}
\includegraphics[width=16cm, angle=0]{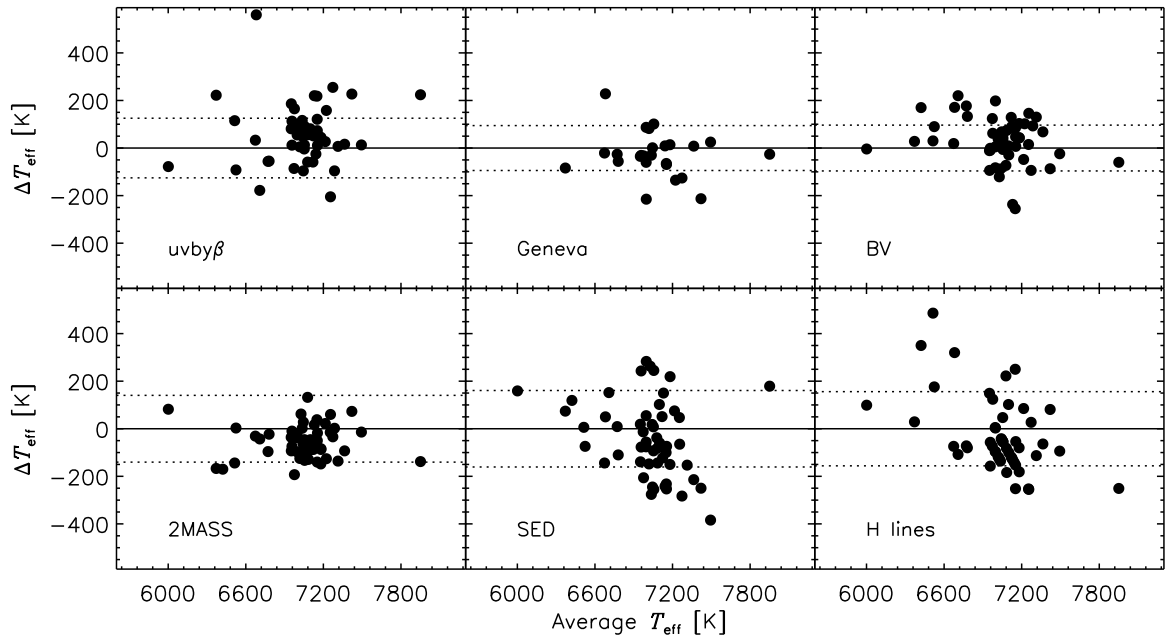}
\caption{The differences between the average photometric effective temperatures and the \teff\ obtained from photometric methods, SED and hydrogen lines. 
The dashed lines show 1-$\sigma$ levels.}
\label{figure2}
\end{figure*}

\subsection{Photometric parameters}

The effective temperatures \teff\ and surface gravities \logg\ of our targets were determined from photometric indices.
These photometric parameters serve as input values for further analysis.
We used $uvby\beta$\ Str\"{o}mgren, Johnson, Geneva and 2MASS photometric data gathered from the General Catalogue of 
Photometric Data \citep{1997A&AS..124..349M} and the 2MASS catalogue \citep{2003tmc..book.....C}.

For 49 stars the atmospheric parameters \teff\ and \logg\ were estimated from the $uvby\beta$ system. 
These parameters were acquired using the method of \citet{1985MNRAS.217..305M}, based on the $V$, $(b-y)$, $m_{1}$, $c_{1}$ and $\beta$ indices.

For 23 stars, Geneva photometry was used to derive the \teff\ and \logg\ values.
The calculations were performed using the \citet{1997A&AS..122...51K} calibration based on the $B2-V1$, $d$ and $m2$ indices. 

Johnson $(B-V)$ colour indices were used to determine the \teff\ of all studied stars.
We applied the $(B-V)$ colour-temperature relation given by \citet{2000AJ....120.1072S}.
For calculations of \teff\ values, \logg\ = 4.0 and solar metallicities were assumed.

Finally, \teff\ values of the stars were derived from the 2MASS photometry \citep{2006A&A...450..735M}, using $(V-K)$ index.
In the calculations, we assumed solar metallicity ($[m/H] = 0.0$) and \logg\ = 4.0 for all the stars.

The results obtained with all these methods are listed in Table\,3.
Uncertainties of the calculated \teff\ and \logg\ were estimated taking into account errors of photometric indices,
reddening (0.02\,mag, as discussed before), metallicity (0.1\,dex), and surface gravity (0.1\,dex), if it was necessary to assume this last parameter.
Finally we derived the average uncertainties of \teff\ and \logg\ for each system (see Table\,3).
The average effective temperatures were calculated using the results from all considered photometric systems.
In Fig.\,\ref{figure2}, these values are compared with individual results for each photometric system.
The dashed lines represent the standard deviations of differences between the average temperatures and values from a given photometric system.
These standard deviations are equal 125, 94, 96 and 140\,K for $uvby\beta$, Geneva, Johnson and 2MASS systems, respectively.
As can be seen in Fig.\,\ref{figure2}, in most cases the obtained effective temperatures are consistent with the average values.
In the case of $uvby\beta$\, the biggest difference was derived for HD\,110379. 
This star is a binary system member, and its photometric colours can be influenced by the light from the second component.

Additionally, the \logg\ values obtained from the $uvby\beta$\ and Geneva indices were compared with each other.
The average \logg\ value for the $uvby\beta$\ system is 4.08\,dex while for the Geneva system, it reaches 4.32\,dex.
As can be inferred from these average values, surface gravities from $uvby\beta$ are in general slightly lower than the Geneva ones.

\subsection{Effective Temperature from SED}

Stellar parameters can be estimated from the SED of a star.
SEDs have to be constructed from spectrophotometry collected in different wavelengths, preferably from ultraviolet until infrared. 
Different parts of SED are sensitive to different stellar parameters.
We used SEDs to obtain \teff\ values, using the code written by Dr. Shulyak (private information).
This code automatically searches for spectrophotometric observations from different data bases.
Several data bases are available with the code, e.g. \citet{1989A&AS...81..221A}, \citet{1976ApJS...32....7B}, \citet{1996BaltA...5..603A}, 
\citet{1985AbaOB..59...83B} and \citet{1992A&AS...92....1G} covering the near-UV, visual, and near-IR wavelengths.
The code can additionally use data from the Space Telescope Imaging Spectrograph \citep*[STIS, Hubble Space Telescope;][]{1998PASP..110.1183W}, 
the International Ultraviolet Explorer \citep*[IUE;][]{2000Ap&SS.273..155W}, and the Ultraviolet Sky Survey Telescope \citep[TD1;][]{1973MNRAS.163..291B,1978csuf.book.....T}.
These archives cover the ultraviolet part of SED. The code allows also to input indices manually, if necessary.

In this study, we generally used the photometric colours of the $uvby\beta$, Geneva, Johnson, and 2MASS systems and the ultraviolet TD1 observations 
as input values. SEDs constructed from these observed spectrophotometric measurements were 
compared with theoretical energy distributions, calculated from the Kurucz's atmospheric models \citep*[ATLAS9 code,][]{1993KurCD..13.....K}.
In these calculations, the solar metallicity ($[m/H]=0$) and the \logg\ value of 4.0\,dex were assumed.
The obtained \teff\ values and their uncertainties are listed in Table\,3. The average error is about 110\,K. 
Differences between the obtained SED \teff\ values and the average photometric values are shown in Fig.\,\ref{figure2}. 
In the figure dashed lines represent standard deviations of 160\,K. As can be seen from Fig.\,\ref{figure2}, the highest difference was 
derived for HD\,209295. This can be caused by the membership of this star to a binary system.

\begin{figure}
\includegraphics[width=8.5cm, angle=0]{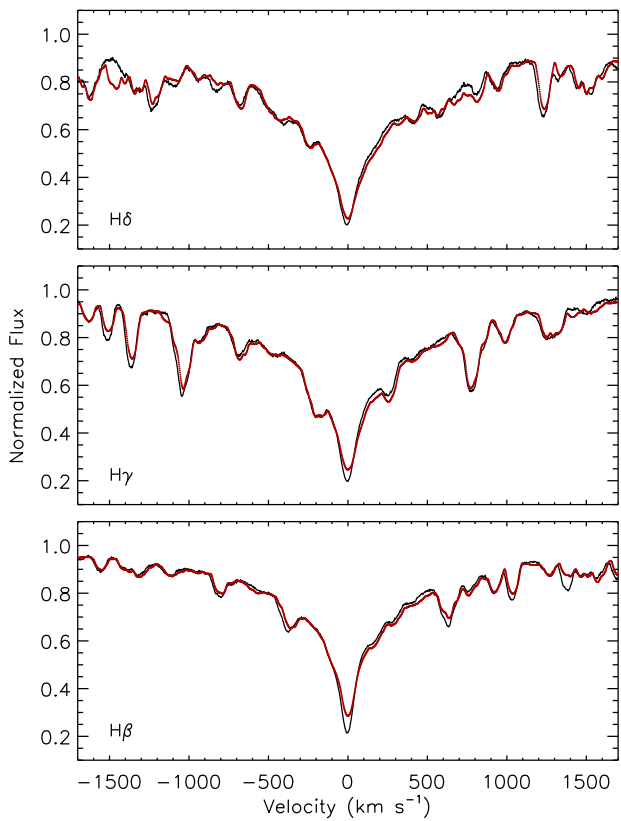}
\caption{The observed Balmer lines (black lines) and the synthetic spectra (red lines) for HD\,23005.}
\label{figure3}
\end{figure}

\begin{figure}
\includegraphics[width=8.5cm, angle=0]{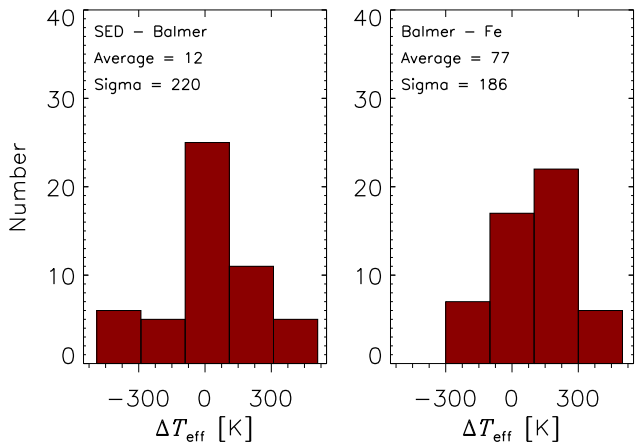}
\caption{The distributions of differences between the effective temperatures determined from the hydrogen lines and SED analysis (left-hand panel) 
and from the hydrogen and Fe lines analysis (right-hand panel).}
\label{figure4}
\end{figure}

\section{Spectroscopic analysis}

In this section, the analysis of high-resolution and high S/N spectra is presented.
Atmospheric parameters were derived from the analysis of hydrogen and metal lines.
All the necessary atmospheric models were calculated with the ATLAS9 code \citep{1993KurCD..13.....K}
that generates hydrostatic, plane-parallel and line-blanketed LTE (local thermodynamic equilibrium) models.
The synthetic spectra were obtained with the SYNTHE code \citep{1981SAOSR.391.....K}.

\subsection{Analysis of hydrogen lines}

The hydrogen lines analysis was applied to obtain the \teff\ values of all stars.
During the analysis of Balmer lines, the \logg\ values were assumed to be 4.0\,dex.
Additionally, the solar metallicity and $v\sin i$ values were fixed during the analysis.
The initial values of $v\sin i$ were taken from the approximate fitting of the synthetic spectra to the observed metal lines.
The analysis was performed taking into account the H$\beta$, H$\gamma$ and H$\delta$ lines.
The method proposed by \citet{2004A&A...425..641C} was applied. 
Initial \teff\ values were taken from previous calculations, including photometric and the SED \teff\ results.
The final effective temperatures were derived minimizing the differences between synthetic and observed spectra.
As an example, the result of the analysis for HD\,23005 is shown in Fig.\,\ref{figure3}.
As can be seen, the observed hydrogen lines fit quite well with the synthetic spectra. 
The small deviations in the core of the lines are caused by the incorrect models, which are not able to explain Balmer line cores. 
The effective temperatures derived from the hydrogen lines and their uncertainties are listed in Table\,4. 
These uncertainties were determined taking into account uncertainties resulting from quality of the spectra (S/N) and assumed values 
of \logg, $[m/H]$ and $v\sin i$. As known, the hydrogen lines are not sensitive to \logg\ in the temperature range of $\gamma$\,Dor stars. 
Because of this, the \logg\ parameter has no significant effect on \teff\ values in our analysis \citep{2002A&A...395..601S, 2005MSAIS...8..130S}.
The obtained uncertainties are in the range of $\sim$ 150-260\,K. 

In Fig.\,\ref{figure2} the obtained values are compared with the average \teff\ calculated from photometric indices.
Standard deviation of these differences is about 200\,K, and is shown by the dashed lines in Fig.\,\ref{figure2}. 
In Fig.\,\ref{figure4} \teff\ parameters from hydrogen lines are compared with the results of SED and iron lines analysis.
The results are consistent within the error bars. 
Standard deviations and average values of these distributions are given in Fig.\,\ref{figure4}.

\begin{figure*}
\includegraphics[width=17cm,angle=0]{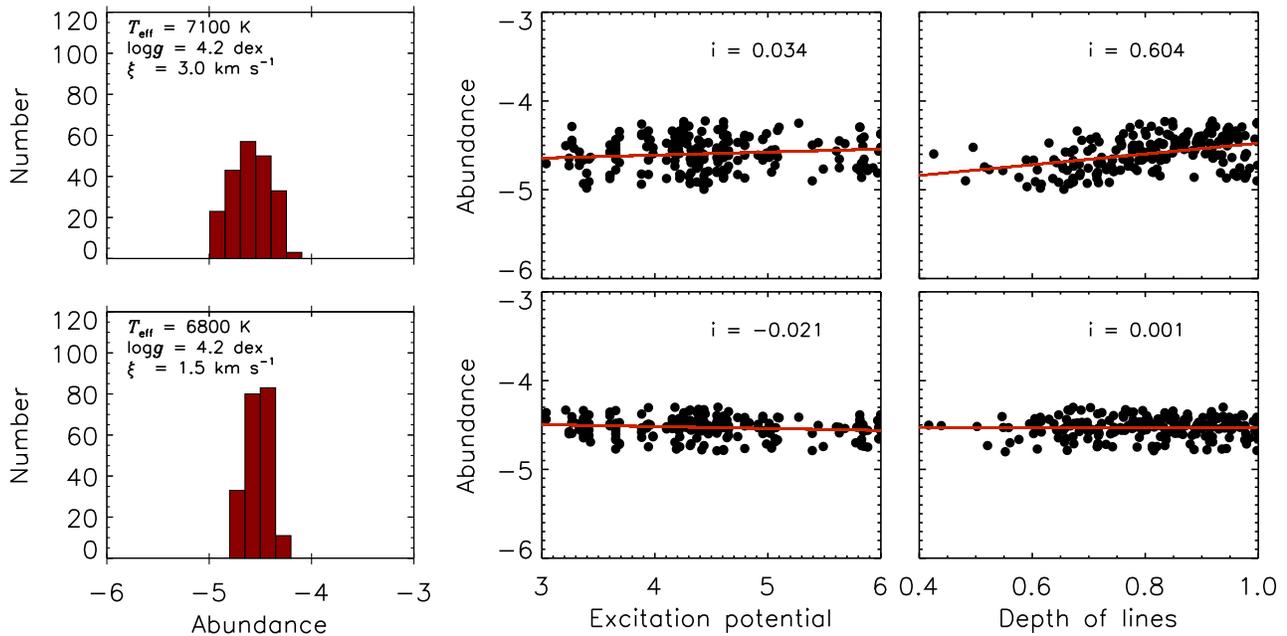}
\caption{The distributions of the derived iron abundances, the \ion{Fe}{i} abundances versus the excitation potential and the lines' depths for 
HD\,126516. The 'i' values illustrate an inclination of the fitted line. The first set of parameters (\teff\ $=7100$\,K, \logg\ $=4.2$, $\xi = 3$\,\kms) 
is incorrect (upper panels), while the second set (6800\,K, 4.2\,dex, 1.5\,\kms) is the right one (lower panels).}
\label{figure5}
\end{figure*}

\subsection{Atmospheric parameters from iron lines analysis}

When \teff\ and \logg\ values were determined from photometric methods, SEDs and hydrogen lines,
the iron lines analysis was performed assuming previously determined parameters as inputs.
The initial analysis of the stellar $v\sin i$ values, revealed slowly and fast rotating stars in our sample.
High rotation velocity causes that most of the spectral lines are blended.
To analyse such spectra and to determine atmospheric parameters (\teff, \logg\ and microturbulent velocity $\xi$ ) 
the spectrum synthesis is the most appropriate method.
To perform our analysis, we followed the same procedure as described in \citet{2015MNRAS.450.2764N}.
The values of \teff, \logg\ and $\xi$ were determined taking into account \ion{Fe}{i} and \ion{Fe}{ii} lines.
\teff\ and $\xi$ parameters are highly sensitive to the strength of the \ion{Fe}{i} lines 
while the \logg\ parameter is almost totally insensitive to it.
The strength of the \ion{Fe}{ii} lines is slightly affected by \teff\ and $\xi$, but depends considerably on the \logg.
Considering the mentioned dependence of the iron lines on atmospheric parameters, we first obtained $\xi$ values 
by looking at the correlation between the \ion{Fe}{i} lines depths and abundances.
Secondly, \teff\ values were derived by checking the correlation between the excitation potential and the abundances calculated from individual \ion{Fe}{i} lines.
In both cases the correlations should be nearly zero, which means that for the proper atmospheric parameters of a star, 
the same iron abundance should be obtained from all iron lines.
The surface gravity values were determined using the ionization balance of the \ion{Fe}{i} and \ion{Fe}{ii} lines.

In Fig.\,\ref{figure5}, we show the distributions of the derived iron abundances (left), excitation potentials versus \ion{Fe}{i} abundances (middle) 
and \ion{Fe}{i} lines depths versus abundances (right) for the star HD\,126516.
Additionally, two sets of the atmospheric parameters for which the iron abundances were calculated are shown.
The upper panels show these relations for the wrong atmospheric parameters, whereas the lower panels demonstrate the right solution.
As we expect that for the correct parameters all iron lines give the same iron abundance within the error bars, 
it is clear that the smallest correlations of line strength, excitation potential, and obtained abundances indicate the correct solution.

The derived values for \teff, \logg\ and $\xi$ together with their uncertainties are given in Table\,4. 
The errors of the analysed parameters were obtained taking into account the effect of other parameters 
on the considered one. The lowest errors of 100\,K for \teff, 0.1\,dex for \logg, and 0.1\,\kms\ for $\xi$
result from the steps adopted in the calculated stellar atmospheric models and synthetic spectra.

\begin{figure*}
\includegraphics[width=17cm, angle=0]{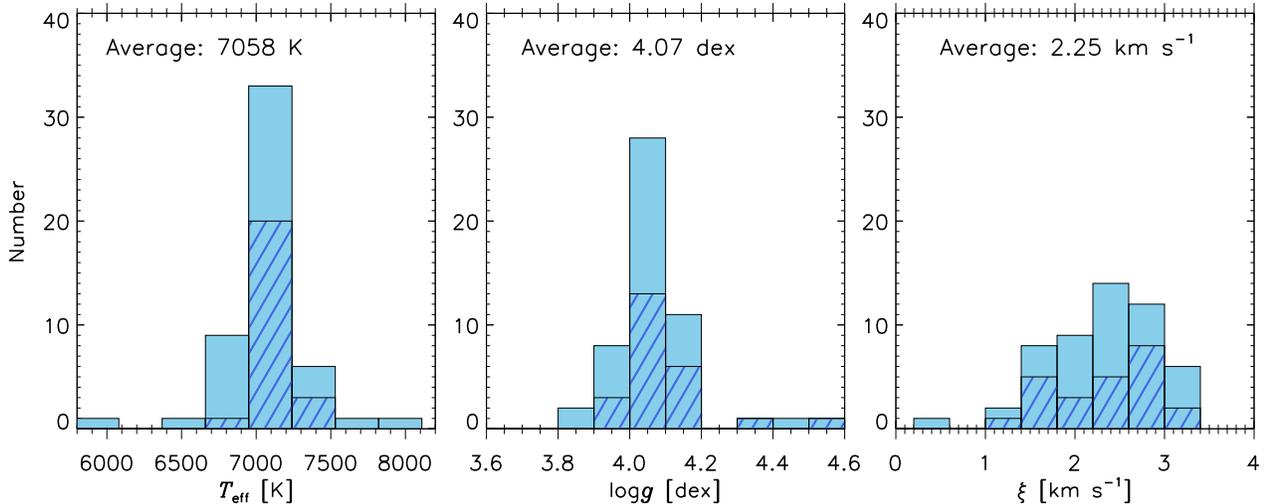}
\caption{The atmospheric parameters distribution of $\gamma$\,Dor stars. Light blue histograms show the distributions for the full sample
while the histograms of the dark blue slant lines illustrates \textit{bona-fide} $\gamma$\,Dor stars.}
\label{figure6}
\end{figure*}

\subsection{Abundances analysis}

After the determination of \teff, \logg\ and $\xi$, chemical abundance analysis was carried out.
In the first step, the spectra of each star were divided into the parts which widths depend mainly on $v\sin i$.
For slowly rotating stars, parts covering only one or a few blended spectral features were selected.
For moderate and fast rotating stars broader spectral ranges, including blends of many lines, were used.
All parts were re-normalised by comparison with theoretical spectra, if necessary.
Then the line identification for selected regions was performed using the line list of Kurucz\footnote{kurucz.harvard.edu/linelists.html} 
\citep{1995KurCD..23.....K} taking into account only these elements that are most important in the given region.
The abundance analysis was carried out by the spectrum synthesis method, which allowed us to determine chemical abundances and $v\sin i$ values at the same time.
We used the classical least square method. Minimum differences between the observed and synthetic spectra indicate the final solution.
After carrying out calculations for each spectral part of a given star, the average values of $v\sin i$ and chemical abundances were derived.
The results are presented in Table\,4 ($v\sin i$ and iron abundances) and Table\,5 (chemical abundances and standard deviations). 

The uncertainties of chemical elements given in Table\,5 are standard deviations. 
The real errors of the elements include the effects of assumptions adopted to build the model of the atmosphere 
and uncertainties of atmospheric parameters (\teff, \logg, $\xi$). 
The assumptions like local thermodynamical equilibrium, plane-parallel geometry,  
and hydrostatic equilibrium were adopted in calculations of atmospheric models and synthetic spectra.
They introduce the error of about 0.1\,dex for calculated chemical abundances \citep{2011mast.conf..314M}. 
Other important factors are used atomic data, analysed wavelength range, quality of the data (resolution, S/N), and normalisation of the spectra. 

To find out the effect of resolution and S/N ratio on the values of the obtained chemical abundances, we selected 
three stars observed by different instruments with different or similar S/N ratios. 
First of these stars (HD\,109799) was observed by FIES (R\,=\,67000) and HARPS (R\,=\,80000) spectrographs. 
Obtained spectra have almost the same S/N ratios ($\sim$ 310). We performed the standard analysis of both spectra and 
obtained a 0.07\,dex difference in iron abundance. 
For the second star (HD\,23005) the spectra were collected by FIES (R\,=\,67000) with S/N\,=\,300 and by HERMES (R\,=\,85000) with 
S/N\,=\,180 ratio. 
According to the S/N ratio and the resolving power of these spectra, both spectra have approximately the same quality. 
In case of this star, we got the difference in iron abundance of about 0.02\,dex. For the third star (HD\,133803) 
we have spectra from HARPS (R\,=\,80000) with S/N\,=\,310 ratio and from HERMES (R\,=\,85000) with S/N\,=\,170 ratio. The spectra 
have nearly the same resolution but different S/N values. When we compared the obtained abundances of iron, 0.13\,dex difference was derived. 
These calculations show, that the resolving power does not have significant influence on abundance determinations in our spectral analysis, 
as all data were taken by high-resolution instruments. On the other hand, the S/N ratio has more important effect.

We also checked the possible influence of quality of the spectrum on atmospheric parameters determination. 
The stars were observed by spectrographs with resolving powers equal 67000, 80000 and 85000. 
Collected spectra have S/N more than 100, often S/N more than 150. 
We can state that the latter value is the recommended one to be used in the abundance analyses, since the improvements obtained with
higher S/N spectra could not be justified by larger investment of observing time.
However, we found no substantial effect of the resolving power and S/N ratio on atmospheric parameters. 
Similar results were also obtained by \citet{2016MNRAS.456.1221R}.

The influence of uncertainties of atmospheric parameters and $v\sin i$ on chemical abundances was examined as well. 
We obtained that 100\,K uncertainty of \teff\ causes the change in element abundance of less than 0.1\,dex. This value increases with increasing \teff. 
On the other hand, the 0.1\,dex error of \logg\ changes the chemical abundances of about 0.04\,dex or less. 
Additionally, we found that for stars in our effective temperature range, 0.1\,\kms\ uncertainty in $\xi$ value changes the element abundance of less than 0.1\,dex. 
The significant influence on the determined abundances can obtain the uncertainty of the $v\sin i$ value.
To examine this effect, we checked the abundance differences caused by changes of $v\sin i$ in the range from $\sim$ 5 to 15 \,\kms, depending on the 
projected rotation velocity of the star. Higher value of $v\sin i$ implies higher value of its uncertainty.
These uncertainties cause differences in abundances ranging from 0.1 to 0.2\,dex. The effect increases with increasing projected rotational velocity. 
Finally, we considered all mentioned uncertainties to calculate the total error of chemical abundances. 
This value can be as high as 0.28\,dex but for most cases it is about 0.20\,dex. 
These errors were also calculated by \cite{2015MNRAS.450.2764N} for hotter stars. They found these values less than 0.20\,dex for their targets. 
The total errors of iron abundances are given in Table\,4.

\subsection{Discussion of the results}

\textit{Effective temperatures}\\

The stellar \teff\ parameters were determined by various methods. The final values were obtained from the iron lines analysis.
As can be seen in the left-hand panel of Fig.\,\ref{figure6}, the \teff\ range for the full sample is between 6000 and 7900\,K, 
while \textit{bona-fide} $\gamma$\,Dor stars have effective temperatures from 7100 to 7300\,K.
This range is in agreement with the results of the previous studies 
\citep{2002MNRAS.333..251H, 2008A&A...478..487B, 2012MNRAS.422.2960T, 2013MNRAS.431.3685T, 2015ApJS..218...27V}.

For 12 stars in our survey, the spectroscopic \teff\ values were determined previously. 
Information about these stars is given in Table\,4. We compared our \teff\ values with the ones given in the literature for these objects.
It turned out that effective temperatures of the mentioned stars are in agreement within an error of 200\,K.
Only in the case of HD\,46304, the difference between \teff\ obtained here and that given by \citet{2013A&A...553A..95M} reaches 400\,K. 
This difference can be explained by the effect of stellar membership in the visual binary system and 
differences in methods of an atmospheric parameter analysis.\\

\begin{figure}
\includegraphics[width=8.5cm,angle=0]{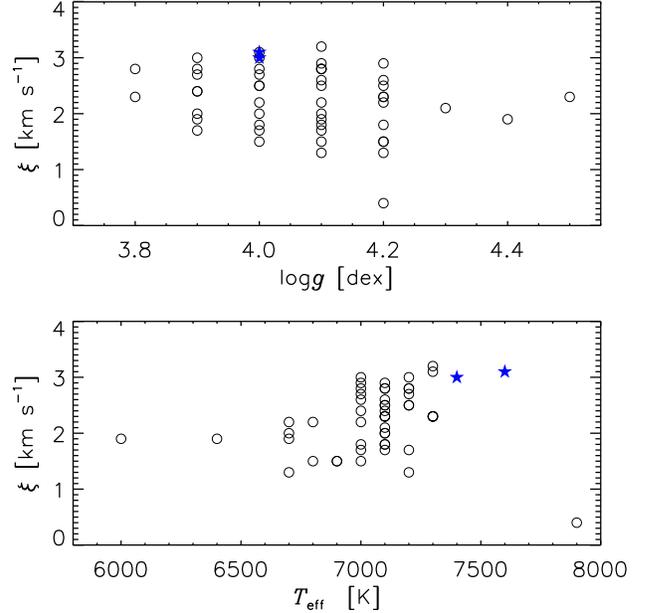}
\caption{The microturbulent velocities as a function of \logg\ and \teff. The star symbols represent Am stars HD\,33204 and HD\,46304.}
\label{figure7}
\end{figure}

\textit{Surface gravities}\\

The \logg\ values were determined from the analysis of the \ion{Fe}{i} and \ion{Fe}{ii} lines (see Table\,4).
The distribution of the obtained \logg\ values is given in the middle panel of Fig.\,\ref{figure6}.
We derived surface gravities between 3.8 and 4.5\,dex. 
\citet{2008A&A...478..487B} obtained a \logg\ range from 3.1 to 4.7 for \textit{bona-fide} and candidate $\gamma$\,Dor stars 
while \citet{2015ApJS..218...27V} found values between 3.3 and 4.5 for a sample of \textit{bona-fide} $\gamma$\,Dor stars only.
In our study, the average \logg\ value amounts to 4.07\,dex for the full sample and the average of 4.09 and 4.05\,dex was found for 
\textit{bona-fide} and candidate $\gamma$\,Dor stars, respectively. 
These values are slightly lower than those given in previous studies (4.16\,dex by \citet{2008A&A...478..487B}; 4.10\,dex by 
\citet{2015ApJS..218...27V}) what indicates that the stars analysed here are more evolved.
From a comparison of our \logg\ values with those found in the literature, we conclude that they agree within 0.2\,dex.\\

\textit{Microturbulent velocities}\\

The obtained microturbulent velocities range from 1.3 to 3.2\,km\,s$^{-1}$ (see right-hand panel of Fig.\,\ref{figure6}) for all stars except for HD\,75202. 
This star is a candidate $\gamma$\,Dor star, listed in a catalogue of contact binary objects \citep{2003CoSka..33...38P}. 
The spectrum of HD\,75202 can be affected by the other system member, which can influence the determined atmospheric parameters.

The range of $\xi$ values is in agreement with the results of \citet{2009A&A...503..973L}, \citet{2014psce.conf..193G}, and \citet{2015MNRAS.450.2764N}.
According to these studies, for effective temperatures between 7000 and 8000\,K $\xi$ values are mostly between 2 and 4\,km\,s$^{-1}$.
The value of this parameter decreases for temperatures lower than $\sim$\,7000\,K and higher than $\sim$\,8000\,K.
In the case of chemically peculiar Am stars, the $\xi$ values are expected to be higher than for normal stars \citep{2009A&A...503..973L}.
We plotted the $\xi$ parameter as a function of \teff\ in the right-hand panel of Fig.\,\ref{figure7}.
As can be seen, the Am stars in our study show the same $\xi$ values as non-chemically peculiar stars.
A similar result for Am stars was obtained by \citet{2015MNRAS.450.2764N} and \citet{2004IAUS..224..131S}.
We also examined the variation of $\xi$ values with surface gravity (left-hand panel of Fig.\,\ref{figure7}). 
The $\xi$ values are lower with increasing values of \logg. The same variations for F type stars were obtained by \citet{2001AJ....121.2159G}.

\citet{2015ApJS..218...27V} found $\xi$ values between 2 and 3.5\,km\,s$^{-1}$ for $\gamma$\,Dor stars, in agreement with our results. 
The small differences are due to the differences in the applied methods and adopted atomic data. \\

\textit{Projected rotational velocities}\\

\begin{figure}
\includegraphics[width=8.5cm, angle=0]{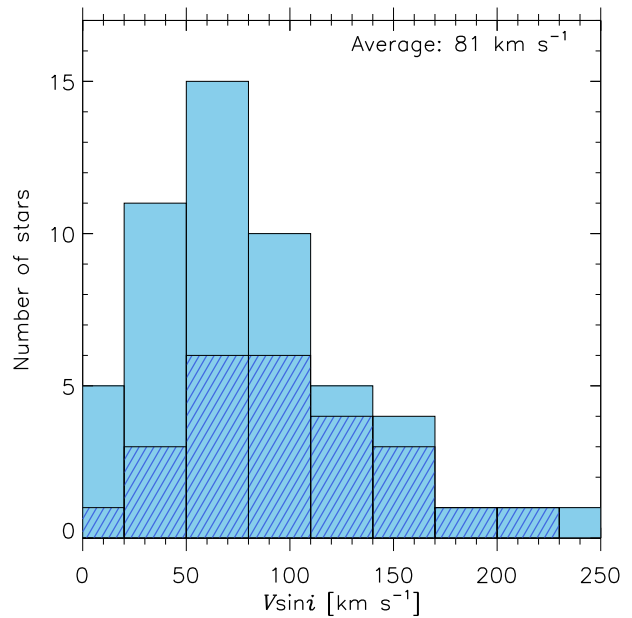}
\caption{The rotational velocity distribution of the stars. The light blue histogram shows the distribution for full sample
while the histogram of dark blue slant lines illustrates the distribution \textit{bona-fide} $\gamma$\,Dor stars.}
\label{figure8}
\end{figure}

The $v\sin i$ values were derived during the analysis of the chemical abundances by the spectrum synthesis method.
The range of the obtained projected rotational velocities is between 5 and 240\,km\,s$^{-1}$.
The distribution of $v\sin i$ is shown in Fig.\,\ref{figure8}.
The average $v\sin i$ value equal to 80\,km\,s$^{-1}$ was calculated taking into account all analysed stars. 
When considering \textit{bona-fide} and candidate $\gamma$\,Dor stars separately, the average values are 97 and 63\,km\,s$^{-1}$, respectively.
In the previous studies, this value obtained for \textit{bona-fide} and candidate $\gamma$\,Dor stars equals 
57\,km\,s$^{-1}$ \citep{2002PASP..114..999H, 2003AJ....126.3058H, 2003AJ....125.2196F, 2004A&A...417..189M, 2006A&A...449..281D, 2015ApJS..218...27V}. 
\citet{2015ApJS..218...27V} gives the range from 12 to 204\,km\,s$^{-1}$ and the average value of 71\,km\,s$^{-1}$ for \textit{bona-fide} 
$\gamma$\,Dor stars. All these values depend on the analysed sample of stars. 
However, both our results and \citet{2015ApJS..218...27V} calculations suggest a great variation of projected 
rotational velocities of \textit{bona-fide} $\gamma$\,Dor stars .

In our study, most stars have high projected rotational velocities ($v\sin i >$\,100\,km\,s$^{-1}$).
However, during the analysis we found some slowly rotating stars (HD\,21788, HD\,104860, HD\,109838 and HD\,126516) with $v\sin i <$\,15\,km\,s$^{-1}$.
Not all chemically peculiar stars from our sample are slowly rotating stars.
HD\,33204 has $v\sin i$ value of 36\,km\,s$^{-1}$ while HD\,46304 has $v\sin i$\,= 242\,km\,s$^{-1}$. 
It has been shown that Am stars generally have smaller rotational velocities than normal stars \citep{1971ApJ...163..333A}.\\

\textit{Chemical abundances}\\

\begin{figure}
\includegraphics[width=8.5cm, angle=0]{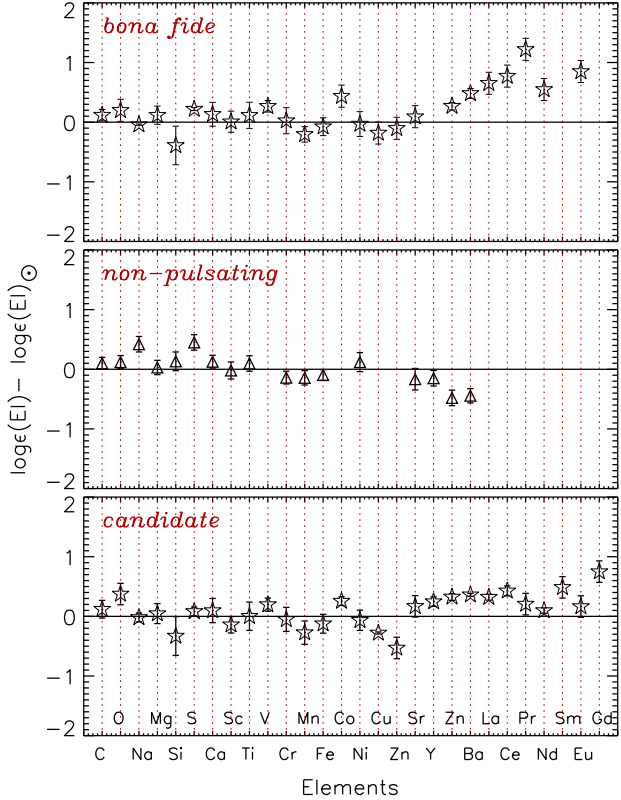}
\caption{Chemical abundances of the \textit{bona-fide}, candidate $\gamma$\,Dor stars and the non-pulsating F type stars. 
Solar abundances and those of non-pulsating stars were taken from \citet{2009ARA&A..47..481A} and \citet{2015MNRAS.450.2764N}, respectively.}
\label{figure9}
\end{figure}

The abundance pattern of $\gamma$\,Dor stars was examined in detail.
The average relative abundance of \textit{bona-fide}, candidate $\gamma$\,Dor, and non-pulsating F type stars were compared.
The abundance distributions of four non-pulsating F type stars were taken from \citet{2015MNRAS.450.2764N}, 
as the same analysis method is used in the current study. This comparison is demonstrated in Fig.\,\ref{figure9}.
As can be seen, the abundances of both \textit{bona-fide} and candidate $\gamma$\,Dor stars are close to the solar abundances.

\begin{figure}
\includegraphics[width=8.5cm, angle=0]{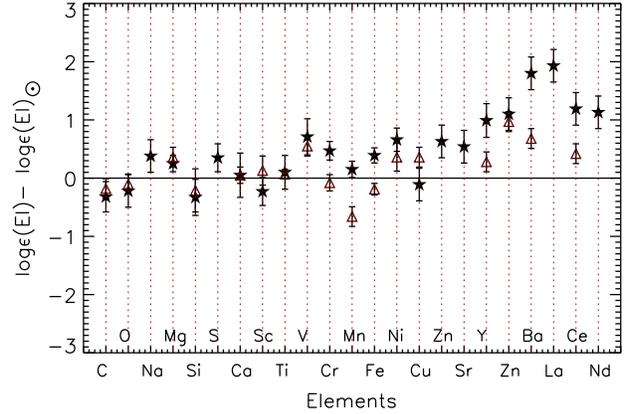}
\caption{Chemical abundances of Am: stars compared with the solar values \citep{2009ARA&A..47..481A}. HD\,33204 is represented by stars and HD\,46304 by triangles.}
\label{figure10}
\end{figure}

The abundances of Am: stars were also examined in detail. As mentioned before, we identified two mild-Am stars, HD\,33204 and HD\,46304. 
We show abundance distributions of these stars in Fig.\,\ref{figure10}. 
A typical Am star exhibits overabundances of iron-peak elements and some heavy elements (Zn, Sr, Zr and Ba), 
but Ca and Sc abundances of these stars are underabundant \citep{2009ssc..book.....G}.
As can be seen in Fig.\,\ref{figure10}, the mild-Am stars in our study have nearly solar abundances of Ca and Sc elements. 
Only HD\,33204 shows overabundances in some heavy elements typical for Am star. In the case of HD\,46304, most of lines are blended 
due to high rotation velocity of the star. These blended lines cause difficulties in abundance calculations.  
The abundance differences between HD\,46304 and a typical Am star can be caused by this effect. 
In the spectral classification process some stars were defined as metal-poor, mostly taking into account Mg, Mn and Fe lines (see Table\,2). 
For these stars, the average abundance of Mg (7.57\,dex) was found to be close to the solar abundance. 
However, the average abundances of Mn (5.13\,dex) and Fe (7.27\,dex) are slightly lower than the solar abundances.
For some of our targets (HD\,26298, HD\,33204, HD\,106103, HD\,110379, and HD\,126516) chemical abundances were already obtained before the present study.
We compared our atmospheric parameters and abundance results with the literature values.
Abundances of HD\,33204 and HD\,10613 were derived by \citet{1999A&A...351..247V} and \citet{2008A&A...483..891F}, respectively.
In their work, similar methods for abundance analysis were used but different atomic data bases were adopted.
For HD\,33204 higher abundances of Sc, Mg and Y were obtained in the present study. Our result is consistent with the Am: type peculiarity of this star.
For HD\,106103 only the Y abundance is different. Abundances of HD\,26298, HD\,110379 and HD\,126516 were derived by \citet{2008A&A...478..487B} 
using the equivalent width method and the different atomic data base. 
In the case of these three stars, the marked differences were detected only for V and Ba abundances.

Additionally, we checked the possible correlations of the element abundances with atmospheric parameters, i.e. \teff, \logg\, and $\xi$. 
Similar as in \citet{2015MNRAS.450.2764N}, no correlations were found.  
Moreover, the element abundances do not depend on $v\sin i$. The same result was found by \citet{2008A&A...483..891F}.
However, \citet{2008JKAS...41...83T} reported negative correlations between $v\sin i$ and C, O and Ca elements.
Finally, we checked the relations between abundances of iron and other elements. 
Strong positive correlations were found for Mg, Si, Ca, Sc, Ti Cr, Ni, Y and Ba. 
Similar correlations between Fe abundances and iron-peak elements were found by \citet{2015MNRAS.450.2764N}.
On the other hand, a negative correlation was obtained between Fe and O abundances.

\section{Discussion and conclusions}

This study presents a detailed analysis of the atmospheric parameters and chemical abundances of a sample \textit{bona-fide} and candidate $\gamma$\,Dor stars. 
We analysed the high-resolution and high S/N spectra of 52 objects. 
The results of the spectral classifications show that the spectral types of $\gamma$\,Dor stars are between A7 and F9 and 
that their luminosity classes range from V to IV. 
During the spectral classifications process, two mild-Am stars, HD\,33204 and HD\,46304 were defined. 
Peculiarities of these stars were checked with the results of the detailed abundance analysis. 
Only for HD\,33204 this peculiarity was confirmed. 
Because of the high rotation velocity of HD\,46304, the peculiarity of this star could not be confirmed. 

To determine the initial atmospheric parameters (\teff\ and \logg), we used photometric indices, SEDs and hydrogen lines. 
The obtained \teff\ values were compared with each other. 
We found that effective temperatures from different methods are mostly in agreement. 
The final atmospheric parameters of the stars were derived from iron lines analysis using the spectrum synthesis method. 
The agreement between the new results obtained with our analysis and those previously available shows the 
robustness of the spectroscopic procedures adopted to analyse the chemical abundances of A-F stars.

For the whole sample, the obtained \teff\ values range from 6000 to 7900\,K, while the obtained \logg\ changes from 3.8 to 4.5\,dex. 
This result corresponds with the obtained luminosity type of the stars.
Additionally, the $\xi$ parameters were derived in the range of 1.3\,-\,3.2\,km\,s$^{-1}$. 
The stars in our sample have mostly moderate and high rotation velocities. The obtained $v\sin i$ values are between 5 and 240\,km\,s$^{-1}$, 
while average values are equal to 97 and 63\,km\,s$^{-1}$ for the \textit{bona-fide} and the candidate stars, respectively. 

\begin{figure}
\includegraphics[width=8.5cm, angle=0]{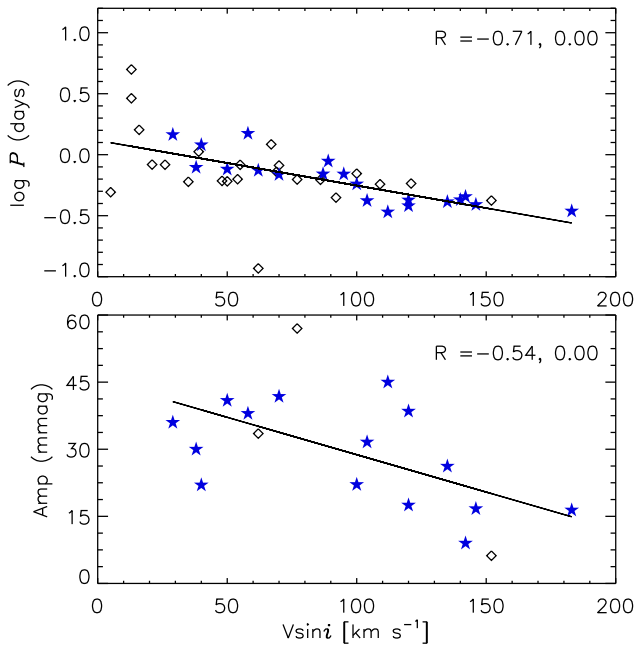}
\caption{The comparison of $v\sin i$ with the pulsation period and amplitude of $\gamma$\,Dor stars. 
\textit{Bona-fide} and candidate $\gamma$\,Dor variables are represented by stars and diamonds, respectively.
The first number of R constant shows strength of the correlation (in the ideal case close to 1) 
while the second number represents deviations of points from the correlations (in the ideal case close to 0).}
\label{figure11}
\end{figure}

\begin{figure}
\includegraphics[width=8.5cm, angle=0]{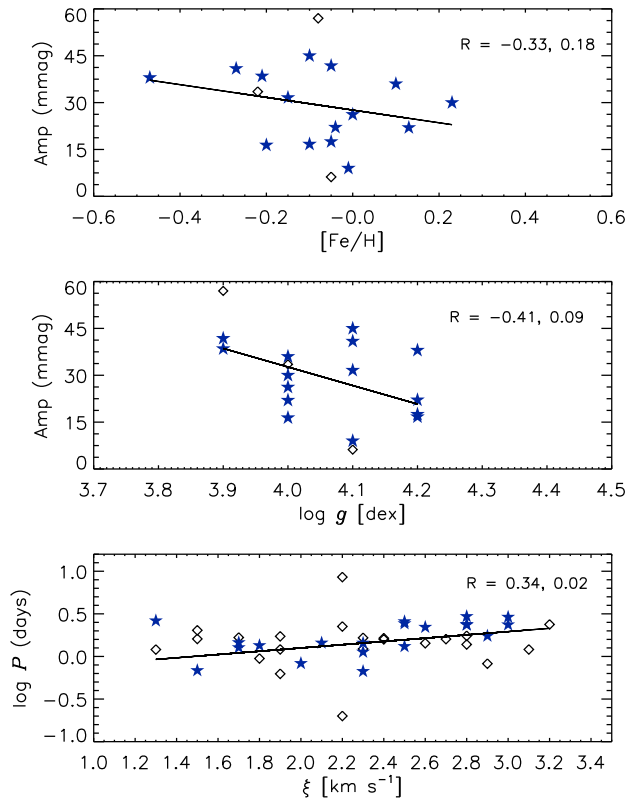}
\caption{The relations between the pulsation period and pulsation amplitude of $\gamma$\,Dor stars and certain parameters given in x-axis. 
Stars and diamonds represent \textit{bona-fide} and candidate $\gamma$\,Dor variables, respectively. 
The R constant has the same meaning as in Fig.\,\ref{figure11}.}
\label{figure12}
\end{figure}

After the determination of accurate stellar parameters, 
relations between \teff, \logg, $\xi$ and $v\sin i$ and pulsation periods and V-band amplitudes of the $\gamma$\,Dor stars were investigated.
The stellar pulsation parameters were taken from the papers given in Table\,2.
The existence of the correlation between the pulsation and rotation periods of variables was suggested by \citet{2011MNRAS.415.3531B}.
Also we found a strong relation between $v\sin i$ parameter and pulsation period, as shown in Fig.\,\ref{figure11}. 
The similar result was obtained in the previous studies \citep[e.g.][]{2013A&A...556A..52T, 2015ApJS..218...27V}. 
This shows that the pulsation periods of stars decrease with increasing $v\sin i$ values. 
This result is in agreement with the theoretical study of \citet{2013MNRAS.429.2500B}, where 
it was shown that g-mode frequencies are shifted to higher frequencies by rotation. 
On the other hand, we could not find a clear correlation between pulsation period and \teff\ values, despite 
the positive relation found by \citet{2015ApJS..218...27V}.
We also found weak correlations between the pulsation amplitude and both \logg\ values and relative iron abundances. 
Additionally, the correlation between pulsation periods and $\xi$ was obtained. 
These correlations are presented in Fig.\,\ref{figure12}. 
As can be seen from this figure, more data is necessary in order to establish the exact relations between those parameters. 

The comprehensive abundance analysis of both \textit{bona-fide} and candidate $\gamma$\,Dor stars was performed using the spectrum synthesis method. 
We compared chemical abundances of \textit{bona-fide} $\gamma$\,Dor stars with those obtained for candidates and F type non-pulsating stars. 
According to these comparisons, no obvious differences were obtained. These stars have abundances close to the solar values \citep{2009ARA&A..47..481A}. 
The derived average iron abundances are equal $7.42$ and $7.38$\,dex for \textit{bona-fide} and candidate $\gamma$\,Dor stars, respectively. 
These values are also close to the solar iron abundances ($7.50$\,dex).

\begin{figure*}
\includegraphics[width=17cm, angle=0]{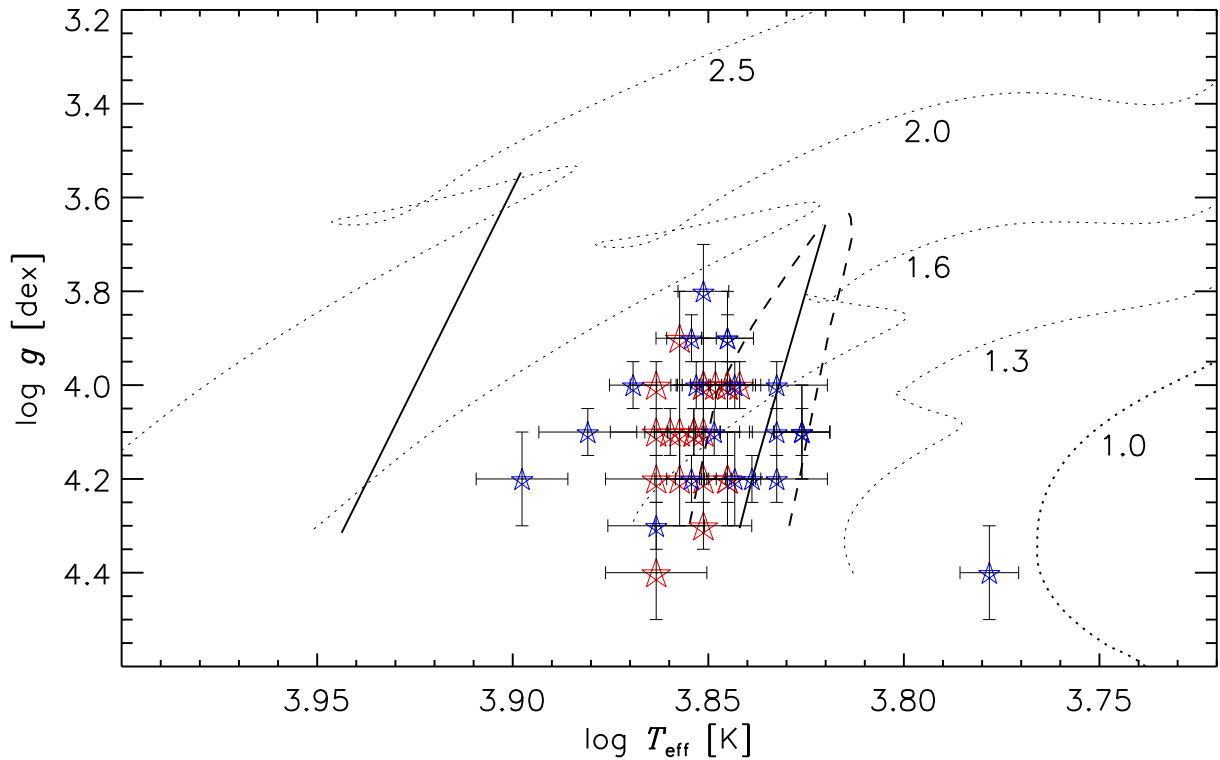}
\caption{Positions of the \textit{bona-fide} (bigger red stars) and candidate (small blue stars) $\gamma$\,Dor 
stars in the theoretical instability strips of the $\gamma$\,Dor (dashed-lines) and $\delta$\,Sct 
(solid lines) stars \citep{2005A&A...435..927D}.}
\label{figure13}
\end{figure*}

In Fig.\,\ref{figure13}, the positions of the studied stars in the theoretical instability strips of $\gamma$\,Dor and $\delta$\,Sct stars are given. 
The evolutionary tracks shown in this figure were calculated with the MESA (Modules for Experiments in Stellar Astrophysics) evolution code 
\citep{2011ApJS..192....3P, 2013ApJS..208....4P, 2015arXiv150603146P}.
All the computed models have an initial hydrogen abundance of X\,=\,0.7, the initial helium abundance of 
Y\,=\,0.28 and use the AGSS09 metal mixture \citep{2009ARA&A..47..481A}. 
The initial metal abundance is Z\,=\,0.02. The OPAL \citep{1996ApJ...464..943I} opacity tables were used. 
All effects of rotation were neglected. The convective zones were determined by the Ledoux criterion. 
For the envelope, we adopted the parameter of the mixing length theory (MLT) of $\alpha_\mathrm{mlt}=2.0$, as the 
theoretical instability strip of $\gamma$\,Dor stars were calculated with this value by \cite{2005A&A...435..927D}. 

The positions of $\gamma$\,Dor stars in both theoretical instability strips have been discussed in the literature. 
In \citet{2011A&A...534A.125U}, $\gamma$\,Dor stars were mostly found outside their theoretical instability strip. 
The same result was also presented in \citet{2010ApJ...713L.192G} and \citet{2012MNRAS.422.2960T, 2013MNRAS.431.3685T}, whereas 
\citet{2013A&A...556A..52T} found them inside the $\gamma$\,Dor instability strip within errors.
As can be seen in Fig.\,\ref{figure13}, \textit{bona-fide} $\gamma$\,Dor stars mostly cluster at the blue edge of the $\gamma$\,Dor instability strip, 
while some of them are located in $\delta$\,Sct domain. Only HD\,104860 is located outside those instability strips. 
Considering the received atmospheric parameters of this star, we conclude that HD\,104860 is not a $\gamma$\,Dor variable.

In this study, we obtained accurate atmospheric parameters and chemical composition of a large sample of the $\gamma$\,Dor stars. 
They are essential in modelling of the pulsation and in the understanding of the real evolutionary status and stellar structure. 
As a result, we found that our stars are mostly located close to the blue edge of the $\gamma$\,Dor instability strip, 
where $\delta$\,Sct pulsation (i.e., pressure modes) is also possible. 
This seems reduce a little the range of the $\gamma$\,Dor pulsation in the classical instability strip described by \citet{2011A&A...534A.125U}.
In the follow-up paper, we plan to perform a detailed spectroscopic study of SB2 $\gamma$\,Dor stars. 
The investigation of the sample of SB2 stars will give us the possibility to examine the probable differences of 
chemical abundances and atmospheric parameters with respect to the single stars studied in this paper.

\section*{Acknowledgements}
The authors would like to thank the reviewer for useful comments and suggestions that helped to improve the publication. 
This work has been partly supported by the Scientific and Technological Research Council of Turkey (TUBITAK) grant numbers 2214-A and 2211-C. 
This article is a part of the Ph.D. thesis of FKA. FKA wishes to express gratitude to professor Dr J. A. Guzik for all assistance given at the start of this project.
EN acknowledges support from the NCN grant No. 2014/13/B/ST9/00902. 
The calculations have been carried out in Wroc{\l}aw Centre for Networking and Supercomputing (http://www.wcss.pl), grant No.\,214.
We thank professor Dr R. O. Gray and Dr. B. Smalley for their helpful comments. 
We are grateful to Dr. D. Shulyak for putting the code for calculating SEDs at our disposal. 
We thank to Dr. G. Catanzaro for putting the code for Balmer lines analysis at our disposal. 
\textquotedblleft Ministerio de Econom\'{\i}a y Competitividad'' (MINECO) and FEDER funds under the \textquotedblleft Ram\'{o}n y Cajal'' subprogram, 
also acknowledges support by the European project SpaceInn (ref. 312844) within the European SPACE program FP7-SPACE-2011-1, 
and from Junta de Andaluc\~{a}(Spanish) local government under project Contribuci\~{a}n Andaluza al proyecto espacial CoRoT 
with reference P12-TIC-2469.  This work is partially based on observations collected at La Silla
Observatory, ESO (Chile) with the FEROS and HARPS spectrographs under
programmes LP178.D-0361, LP182.D-0356, and LP185.D-0056. EP and MR acknowledge 
financial support from the FP7 project \textquotedblleft SpaceInn: Exploitation of Space Data for Innovative Helio and Asteroseismology" 
from PRIN-INAF 2014 {\it Galactic Archaelogy}. This research has made use of the SIMBAD data base, operated at CDS, 
Strasbourq, France.


\newpage

\begin{table*}
\centering
  \caption{Spectroscopic observations and the spectral clasification of the investigated stars.}
  \begin{tabular}{@{}llccllcc@{}}
  \hline
     HD  &  Instruments  &Number of & V      &    Sp type  &    Sp type      & Notes &References  \\
   number&               &spectra   &(mag)   & (Simbad)    & (this study)    &       &            \\
 \hline
 009365**   & FIES 	&1 & 8.23 & F0      	&F1\,V & $\gamma$\,Dor& 1\\
 019655   & FIES 	&3 & 8.62 & F2\,V    	&F1\,V\,nn &  $\gamma$\,Dor&3\\
 021788   & FIES 	&2 & 7.50 & F0     	&F3\,V & cand $\gamma$\,Dor&2\\
 022702   & FIES 	&3 & 8.80 & A2     	&F1\,IV & $\gamma$\,Dor& 3\\
 023005   & FIES 	&4 & 5.82 & F0\,IV   	&F1\,IV\,nn& cand $\gamma$\,Dor& 2\\
 023585   & FIES 	&1 & 8.36 & F0\,V    	&F0\,V&  $\gamma$\,Dor&3\\
 026298**   & FIES 	&1 & 8.16 & F1\,V    	&F2\,V& cand $\gamma$\,Dor& 4\\
 033204   & FIES 	&1 & 6.01 & A5\,m    	&A7\,V Am: & cand $\gamma$\,Dor &5\\
 046304   & FIES 	&3 & 5.60 & F0\,V    	&A8\,V Am: &cand $\gamma$\,Dor &16\\
 063436   & FIES 	&1 & 7.46 & F2     	&F0\,IV&  $\gamma$\,Dor &6\\
 089781   & FIES 	&1 & 7.48 & F0    	&F1\,V &  $\gamma$\,Dor &1\\
 099267   & FIES 	&3 & 6.87 & F0     	&F1\,V &  $\gamma$\,Dor&6\\
 099329   & FIES 	&3 & 6.37 & F3\,IV  	&F2\,IV\,nn & $\gamma$\,Dor&1\\
 104860   & FIES 	&2 & 7.91 & F8     	&G0\,/\,F9\,V& cand $\gamma$\,Dor&2\\
 106103   & FIES	&1 & 8.12 & F5\,V   	&F2\,-\,3\,V & cand $\gamma$\,Dor& 14 \\
 107192   & FIES 	&1 & 6.28 & F2\,V   	&F1\,IV& cand $\gamma$\,Dor&7\\
 109032   & FIES 	&5 & 8.09 & F0     	&F1\,V& cand $\gamma$\,Dor&2\\
 109799   & FIES 	&1 & 5.45 & F1\,IV   	&F2\,IV& cand $\gamma$\,Dor&2\\
 109838   & FIES 	&1 & 8.04 & F2\,V    	&F2\,IV& cand $\gamma$\,Dor&2\\
 110379   & FIES 	&1 & 3.44 & F0\,IV   	&F1\,-\,F2\,V& cand $\gamma$\,Dor&4\\
 112429   & FIES 	&1 & 5.24 & F0\,IV-V 	&F3\,IV&  $\gamma$\,Dor&6\\
 118388   & FIES 	&3 & 7.98 & F2     	&F5\,V\,m\,-\,3& cand $\gamma$\,Dor&8\\
 126516**   & FIES 	&2 & 8.31 & F3\,V    	&F5\,V& cand $\gamma$\,Dor&4\\
 130173**   & FIES 	&1 & 6.88 & F3\,V    	&F5\,V\,m\,-\,3& cand $\gamma$\,Dor&9\\
 155154   & FIES 	&3 & 6.18 & F0\,IV-Vn	&F2\,IV\,nn& $\gamma$\,Dor&10\\
 165645   & FIES 	&3 & 6.36 & F0\,V    	&F1\,V\,nn& cand $\gamma$\,Dor&6\\
 169577   & FIES 	&5 & 8.65 & F0     	&F1V&  $\gamma$\,Dor& 11\\
 187353   & FIES 	&3 & 7.55 & F0     	&F1\,IV/V& cand $\gamma$\,Dor&2\\
 206043   & FIES 	&1 & 5.87 & F2\,V    	&F1\,V\,nn&  $\gamma$\,Dor& 10\\
 075202   &HARPS 	&5 & 7.75 & A3\,IV      &A7\,V & cand $\gamma$\,Dor& 8\\
 091201   &HARPS 	&5 & 8.12 &F1\,V/IV  	&F1\,V\,/\,IV& cand $\gamma$\,Dor&2\\
 103257   &HARPS 	&5 & 6.62 &F2\,V     	&F2\,V\,m\,-\,2& cand $\gamma$\,Dor&2\\
 113357   &HARPS 	&14& 7.87 &F0\,V     	&F2\,V\,m\,-\,2& cand $\gamma$\,Dor&2\\
 133803   &HARPS 	&4 & 8.15 &A9\,V     	&F2\,IV\,m\,-\,2& cand $\gamma$\,Dor&2\\
 137785   &HARPS 	&6 & 6.43 &F2\,V        &F2\,V& cand  $\gamma$\,Dor&2\\
 149989   &HARPS 	&6 & 6.29 &A9\,V        &F1\,V\,nn\,m-4& $\gamma$\,Dor&4\\
 188032   &HARPS        &10& 8.14 &A9\,/\,F0\,V &A9\,V& cand $\gamma$\,Dor&2\\
 197451**   &HARPS        &3 & 7.18 &F1           &F0\,V& cand $\gamma$\,Dor&2\\
 206481   &HARPS 	&7 & 7.86 &F0\,V     	&F2\,V&  $\gamma$\,Dor&2\\
 224288   &HARPS 	&5 & 8.04 &F0\,V     	&F2\,IV/V& cand $\gamma$\,Dor&2\\
 112934   &HERCULES	&2 & 6.57 &A9\,V     	&A9\,V& cand $\gamma$\,Dor&4\\
 115466   &HERCULES	&2 & 6.89 &F0     	&F1\,IV/V& $\gamma$\,Dor&12\\
 124248   &HERCULES	&2 & 7.17 &A8\,V    	&A8\,-\,A7\,V&  $\gamma$\,Dor&12\\
 171834   &HERCULES	&4 & 5.45 &F3\,V     	&F3\,V& $\gamma$\,Dor&15\\
 172416   &HERCULES	&23& 6.62 &F5\,V     	&F6\,V& cand $\gamma$\,Dor&2\\
 175337   &HERCULES	&2 & 7.36 &F5\,V     	&F2\,V&  $\gamma$\,Dor&12\\
 187028   &HERCULES	&2 & 7.60 &F0\,V     	&F2\,V &  $\gamma$\,Dor&4\\
 209295**   &HERCULES	&2 & 7.32 &A9\,/\,F0\,V &A9\,/\,F0\,V\,-\,IV&  $\gamma$\,Dor&4\\
 216910   &HERCULES	&2 & 6.69 &F2\,IV    	&F2\,V&  $\gamma$\,Dor&4\\
 224638   &HERCULES	&2 & 7.48 &F0      	&F2\,-\,F3\,IV&  $\gamma$\,Dor&6\\
 224945   &HERCULES	&1 & 6.62 &A3     	&A9\,V\,/\,IV&  $\gamma$\,Dor&6\\
 041448   &HERMES  	&1 & 7.62 & A9\,V 	& A9\,V &  $\gamma$\,Dor&6\\
\hline
\end{tabular}

\begin{description}
 \item[ ] References - (1) \citet{2007AJ....133.1421H}; (2) \citet{1999MNRAS.309L..19H}; (3) \citet{2000A&A...358..287M}; (4) \citet{2006A&A...449..281D}
 \item[ ] (5) \cite{eyerphd}; (6) \cite{2011AJ....142...39H}; (7) \cite{1998A&A...337..790A}; (8) \cite{2011MNRAS.414.2602D}
 \item[ ] (9) \cite{2003AJ....125.2196F}; (10) \cite{2001AJ....122.3383H}; (11)\cite{2003A&A...406..203P}; (12) \cite{2005AJ....129.2026H}
 \item[ ] (13) \cite{2002MNRAS.333..251H}; (14) \cite{1995IBVS.4195....1K}; (15) \cite{2011arXiv1111.1840U}; (16)\cite{mathias2003}
 \item[ ] Notations : IV\,/\,V=between IV\,-\,V, IV-V\,=\,whether IV or V, nn=very rapid rotators, m-* = metallicity class where * represents 
 \item[ ] number, ``Am:'' defines a mild Am star, cand\,=\,candidate, **=SB1 stars.
 \end{description}

\end{table*}

\begin{table*}
\centering
\footnotesize
  \caption{The $E(B-V)$ values and atmospheric parameters derived from the photometric indices and SED analysis.}
  \begin{tabular}{@{}ccccccccccc@{}}
  \hline         
     HD    &   $E(B-V)$        & $E(B-V)$             &  \teff\        &\logg\               &  \teff\            & \logg\             &\teff\              & \teff\         & \textbf{ \teff\ }          &\teff\ \\
   number  &    $^{Map}$       & $^{NaD_2}$           & $^{uvby\beta}$ &  $^{uvby\beta}$     &    $^{Geneva}$     &  $^{Geneva}$       &    $^{UBV}$        &   $^{2MASS}$   & \textbf{${Average}^*$}     &$^{SED}$          \\
           &    (mag)          &    (mag)             & (K)            &   (dex)             &    (K)             & (dex)              & (K)                & (K)            &  \textbf{    (K)  }        &(K)          \\  
           &                   &                      &  $\pm$\ 95     & $\pm$\ 0.10         & $\pm$\ 125         & $\pm$\ 0.11        & $\pm$\ 170         & $\pm$\ 80      &          $\pm$\ 245        &\\
 \hline 
 009365    & 0.01 & 0.00 & 7050 & 4.05 &        &      & 7200 & 6940 & 7060 & 7280\,\,$\pm$\,\,190 \\
 019655    & 0.03 & 0.03 & 6950 & 3.96 & 6850   & 4.06 & 7110 & 7130 & 7010 & 6800\,$\pm$\,150 \\
 021788    & 0.02 & 0.02 & 6530 & 3.46 &        &      & 6930 & 6680 & 6750 & 6860\,$\pm$\,200 \\
 022702    & 0.03 & 0.03 &      &      & 6940   & 4.28 & 7050 & 7040 & 7010 & 6800\,$\pm$\,100 \\
 023005    & 0.00 & 0.00 & 7030 & 3.88 & 6920   & 4.07 & 6940 & 6870 & 6940 & 6970\,$\pm$\,120 \\
 023585    & 0.03 & 0.00 & 7530 & 4.31 & 7080   & 4.26 & 7180 & 7220 & 7250 & 6990\,$\pm$\,170 \\
 026298    & 0.01 & 0.00 & 6730 & 4.11 & 6720   & 4.38 & 6910 & 6820 & 6790 & 6670\,$\pm$\,130 \\
 033204    & 0.00 & 0.00 & 7650 & 4.11 & 7210   & 4.04 & 7330 & 7510 & 7425 & 7170\,$\pm$\,150 \\
 046304    & 0.00 & 0.00 & 7380 & 3.88 & 7370   & 4.25 & 7430 & 7270 & 7360 & 7150\,$\pm$\,150 \\
 063436    & 0.00 & 0.00 & 7350 & 4.44 &        &      & 6890 & 7090 & 7110 & 7280\,$\pm$\,110 \\
 089781    & 0.04 & 0.00 & 7090 & 4.03 &        &      & 7050 & 7180 & 7110 & 7060\,$\pm$\,130 \\
 099267    & 0.00 & 0.00 & 7050 & 4.01 &        &      & 7110 & 7030 & 7060 & 7060\,$\pm$\,100 \\
 099329    & 0.00 & 0.00 & 7070 & 4.02 & 6940   & 4.23 & 7000 & 6940 & 6990 & 6870\,$\pm$\,100 \\
 104860    & 0.03 & 0.00 & 5920 & 4.65 &        &      & 6000 & 5970 & 5960 & 6160\,$\pm$\,110\\
 106103    & 0.00 & 0.00 & 6710 & 4.45 & 6650   & 4.55 & 6690 & 6590 & 6660 & 6530\,$\pm$\,100 \\
 107192    & 0.00 & 0.01 & 7090 & 4.26 & 7010   & 4.46 & 6910 & 6830 & 6960 & 7050\,$\pm$\,160\\
 109032    & 0.00 & 0.00 & 7180 & 4.31 &        &      & 7070 & 7030 & 7090 & 7040\,$\pm$\,120\\
 109799    & 0.00 & 0.00 & 7020 & 4.07 & 6940   & 4.33 & 7060 & 6830 & 6960 & 6870\,$\pm$\,150\\
 109838    & 0.02 & 0.00 & 7060 & 4.12 &        &      & 7250 & 7170 & 7160 & 7000\,$\pm$\,250\\
 110379    & 0.00 & 0.00 & 7240 & 4.06 &        &      & 6850 & 5720 & 6600 & 6730\,$\pm$\,300\\
 112429    & 0.00 & 0.00 & 7210 & 4.20 & 7200   & 4.40 & 7280 & 7030 & 7180 & 7010\,$\pm$\,100\\
 118388    & 0.01 & 0.01 &      &      &        &      & 6590 & 6230 & 6410 & 6540\,$\pm$\,250\\
 126516    & 0.01 & 0.00 & 6630 & 4.38 &        &      & 6540 & 6350 & 6510 & 6520 \,$\pm$\,250\\
 130173    & 0.00 & 0.01 & 6430 & 3.77 &        &      & 6610 & 6450 & 6500 & 6570\,$\pm$\,200\\
 155154    & 0.00 & 0.00 & 7170 & 4.04 &        &      & 7160 & 7130 & 7150 & 7080\,$\pm$\,140\\
 165645    & 0.00 & 0.00 & 7320 & 4.02 &        &      & 7440 & 7220 & 7330 & 7160\,$\pm$\,160\\
 169577    & 0.15 & 0.03 & 7050 & 4.24 &        &      & 7400 & 7350 & 7270 & 7190\,$\pm$\,300\\
 187353    & 0.00 & 0.03 & 7020 & 4.10 &        &      & 7000 & 7040 & 7020 & 7040\,$\pm$\,300\\
 206043    & 0.00 & 0.00 & 7180 & 4.05 &        &      & 7110 & 6830 & 7040 & 7200\,$\pm$\,120\\
 075202    & 0.00 & 0.01 & 8180 & 4.06 & 7840   & 4.21 & 7890 & 7680 & 7900 & 8130\,$\pm$\,250\\
 091201    & 0.01 & 0.01 & 7070 & 4.10 &        &      & 7090 & 6980 & 7050 & 6960\,$\pm$\,350\\
 103257    & 0.00 & 0.00 & 6890 & 3.90 &        &      & 7100 & 6940 & 6980 & 6960\,$\pm$\,100\\
 113357    & 0.01 & 0.01 & 7150 & 4.26 &        &      & 7100 & 6900 & 7050 & 6930\,$\pm$\,300\\
 133803    & 0.01 & 0.02 & 7140 & 4.12 &        &      & 6940 & 7030 & 7040 & 7000\,$\pm$\,150\\
 137785    & 0.00 & 0.00 & 7110 & 4.06 &        &      & 7050 & 6820 & 7000 & 6900\,$\pm$\,250\\
 149989    & 0.00 & 0.00 & 7180 & 4.08 & 7070   & 4.43 & 7210 & 7040 & 7120 & 7000\,$\pm$\,100\\
 188032    & 0.00 & 0.00 & 7230 & 4.20 & 7080   & 4.47 & 7200 & 7130 & 7160 & 6900\,$\pm$\,200\\
 197451    & 0.01 & 0.02 & 7370 & 3.93 &        &      & 6900 & 7130 & 7130 & 7050\,$\pm$\,300\\
 206481    & 0.00 & 0.00 & 7150 & 4.24 & 7010   & 4.52 & 6950 & 7040 & 7040 & 6760\,$\pm$\,150\\
 224288    & 0.00 & 0.00 & 7140 & 4.22 & 6940   & 4.40 & 7040 & 6810 & 6980 & 6770\,$\pm$\,120\\
 112934    & 0.00 & 0.00 & 7120 & 4.14 & 7150   & 4.56 & 7220 & 7080 & 7140 & 6900\,$\pm$\,160\\
 115466    & 0.00 & 0.00 & 6970 & 3.93 &        &      & 6960 & 6940 & 6960 & 7200\,$\pm$\,130\\
 124248    & 0.00 & 0.00 & 7220 & 4.16 &        &      & 7220 & 7100 & 7180 & 7400\,$\pm$\,130\\
 171834    & 0.00 & 0.00 & 6720 & 4.03 & 6750   & 4.37 & 6950 & 6680 & 6780 & 6780\,$\pm$\,200\\
 172416    & 0.00 & 0.00 & 6590 & 4.13 & 6290   & 3.68 & 6400 & 6200 & 6370 & 6445\,$\pm$\,100 \\
 175337    & 0.00 & 0.00 & 7090 & 4.14 &        &      & 6900 & 7090 & 7030 & 7290\,$\pm$\,160\\
 187028    & 0.00 & 0.00 & 7270 & 4.34 & 7090   & 4.47 & 7240 & 7010 & 7150 & 6920\,$\pm$\,150\\
 209295    & 0.00 & 0.00 & 7510 & 4.97 & 7470   & 4.25 & 7470 & 7480 & 7480 & 7110\,$\pm$\,220\\
 216910    & 0.00 & 0.00 & 7070 & 4.07 & 6930   & 4.27 & 6950 & 6880 & 6960 & 7390\,$\pm$\,150\\
 224638    & 0.00 & 0.00 & 7140 & 4.06 &        &      & 7160 & 6960 & 7090 & 6940\,$\pm$\,200\\
 224945    & 0.00 & 0.00 &      &      &        &      & 7268 & 7238 & 7250 & 7300\,$\pm$\,160\\
 041448    & 0.00 & 0.00 & 7240 & 4.13 &        &      & 7170 & 7180 & 7200 & 7290\,$\pm$\,150 \\
  \hline
\end{tabular}

\begin{description}
 \item[*] Represents the average values of effective temperature obtained from different photometric systems. 
\end{description}

\end{table*} 

\begin{table*}
\centering
  \caption{Atmospheric parameters obtained from the hydrogen and iron lines analysis.}
  \begin{tabular}{@{}llccccc@{}}
  \hline
     HD   & \teff\ $^{H lines}$ & \teff\ $^{Fe lines}$&  \logg\ $^{Fe lines}$  &  $\xi$ & $v\sin i$     & $\log \epsilon$ (Fe) \\
   number &	(K)             &     (K)	      &	 (dex)                 &  (km/s)&	(km/s)  &   (dex)	  \\
 \hline   
 009365    & 7000\,$\pm$\,170  & 7200\,$\pm$\,100 & 3.9\,$\pm$\,0.1 & 2.7\,$\pm$\,0.2 & 77\,$\pm$\,1  & 7.39\,$\pm$\,0.22 \\
 019655    & 7000\,$\pm$\,210  & 7100\,$\pm$\,100 & 4.1\,$\pm$\,0.3 & 2.8\,$\pm$\,0.4 & 222\,$\pm$\,5 & 7.32\,$\pm$\,0.23 \\
 021788    & 6600\,$\pm$\,140  & 6700\,$\pm$\,100 & 4.1\,$\pm$\,0.2 & 2.2\,$\pm$\,0.1 & 13\,$\pm$\,1  & 7.26\,$\pm$\,0.21 \\
 022702    & 7000\,$\pm$\,190 & 7200\,$\pm$\,200 & 4.2\,$\pm$\,0.2 & 2.5\,$\pm$\,0.3 & 146\,$\pm$\,2 & 7.40\,$\pm$\,0.24\\
 023005    & 7100\,$\pm$\,150 & 7000\,$\pm$\,100 & 3.9\,$\pm$\,0.1 & 2.4\,$\pm$\,0.1 & 48\,$\pm$\,1  & 7.61\,$\pm$\,0.21 \\
 023585*   & 7300\,$\pm$\,250 & 7200\,$\pm$\,200 & 4.1\,$\pm$\,0.2 & 2.8\,$\pm$\,0.3 & 113\,$\pm$\,3 & 7.40\,$\pm$\,0.24 \\
 026298*   & 6700\,$\pm$\,150 & 6700\,$\pm$\,100 & 4.1\,$\pm$\,0.1 & 2.0\,$\pm$\,0.2 & 53\,$\pm$\,2  & 7.20\,$\pm$\,0.22\\
 033204*   & 7500\,$\pm$\,230 & 7600\,$\pm$\,200 & 4.0\,$\pm$\,0.1 & 3.1\,$\pm$\,0.1 & 36\,$\pm$\,2  & 7.97\,$\pm$\,0.26\\
 046304*   & 7300\,$\pm$\,260 & 7400\,$\pm$\,100 & 4.0\,$\pm$\,0.1 & 3.0\,$\pm$\,0.4 & 242\,$\pm$\,12& 7.31\,$\pm$\,0.27\\
 063436*   & 7000\,$\pm$\,170 & 7000\,$\pm$\,100 & 3.9\,$\pm$\,0.1 & 1.7\,$\pm$\,0.2 & 70\,$\pm$\,1  & 7.45\,$\pm$\,0.22\\
 089781    & 7000\,$\pm$\,180 & 7200\,$\pm$\,100 & 4.2\,$\pm$\,0.2 & 1.3\,$\pm$\,0.2 & 120\,$\pm$\,3 & 7.45\,$\pm$\,0.23 \\
 099267    & 7000\,$\pm$\,170 & 7000\,$\pm$\,100 & 4.2\,$\pm$\,0.2 & 2.9\,$\pm$\,0.3 & 100\,$\pm$\,2 & 7.46\,$\pm$\,0.23 \\
 099329*   & 6900\,$\pm$\,200 & 7100\,$\pm$\,200 & 4.1\,$\pm$\,0.2 & 2.6\,$\pm$\,0.3 & 142\,$\pm$\,2 & 7.49\,$\pm$\,0.24 \\
 104860    & 6100\,$\pm$\,140 & 6000\,$\pm$\,100 & 4.4\,$\pm$\,0.2 & 1.9\,$\pm$\,0.1 & 16\,$\pm$\,2 & 7.34\,$\pm$\,0.21\\
 106103*   & 6600\,$\pm$\,150 & 6700\,$\pm$\,100 & 4.1\,$\pm$\,0.2 & 1.3\,$\pm$\,0.1 & 21\,$\pm$\,1  & 7.40\,$\pm$\,0.21\\
 107192    & 6900\,$\pm$\,160 & 7000\,$\pm$\,100 & 3.9\,$\pm$\,0.2 & 2.8\,$\pm$\,0.2 & 69\,$\pm$\,1  & 7.32\,$\pm$\,0.22\\
 109032    & 7000\,$\pm$\,170 & 7000\,$\pm$\,100 & 4.2\,$\pm$\,0.2 & 2.6\,$\pm$\,0.2 & 100\,$\pm$\,1 & 7.42\,$\pm$\,0.22\\
 109799*   & 6900\,$\pm$\,140 & 7000\,$\pm$\,100 & 4.0\,$\pm$\,0.1 & 1.8\,$\pm$\,0.1 & 39\,$\pm$\,2 & 7.51\,$\pm$\,0.21\\
 109838    & 7000\,$\pm$\,140 & 6900\,$\pm$\,100 & 4.2\,$\pm$\,0.1 & 1.5\,$\pm$\,0.1 & 13\,$\pm$\,1  & 7.46\,$\pm$\,0.21\\
 110379*   & 7000\,$\pm$\,150 & 7100\,$\pm$\,100 & 4.1\,$\pm$\,0.1 & 1.8\,$\pm$\,0.2 & 34\,$\pm$\,6 & 7.37\,$\pm$\,0.21\\
 112429    & 7100\,$\pm$\,170  & 7200\,$\pm$\,100 & 3.9\,$\pm$\,0.2 & 3.0\,$\pm$\,0.2 & 120\,$\pm$\,3 & 7.29\,$\pm$\,0.23\\
 118388    & 6800\,$\pm$\,170 & 6700\,$\pm$\,100 & 4.1\,$\pm$\,0.2 & 1.9\,$\pm$\,0.2 & 121\,$\pm$\,8 & 7.27\,$\pm$\,0.22\\
 126516*   & 7000\,$\pm$\,140 & 6800\,$\pm$\,200 & 4.2\,$\pm$\,0.2 & 1.5\,$\pm$\,0.2 & 5\,$\pm$\,1   & 7.50\,$\pm$\,0.23\\
 130173    & 6700\,$\pm$\,160 & 6800\,$\pm$\,200 & 4.0\,$\pm$\,0.2 & 2.2\,$\pm$\,0.2 & 62\,$\pm$\,3  & 7.28\,$\pm$\,0.23\\
 155154    & 7100\,$\pm$\,200  & 7000\,$\pm$\,100 & 4.0\,$\pm$\,0.2 & 3.0\,$\pm$\,0.3 & 183\,$\pm$\,6 & 7.30\,$\pm$\,0.22\\
 165645    & 7200\,$\pm$\,180  & 7300\,$\pm$\,200 & 4.1\,$\pm$\,0.1 & 3.2\,$\pm$\,0.2 & 152\,$\pm$\,4 & 7.36\,$\pm$\,0.28\\
 169577    & 7000\,$\pm$\,160 & 7100\,$\pm$\,200 & 4.2\,$\pm$\,0.1 & 1.8\,$\pm$\,0.2 & 62\,$\pm$\,4  & 7.79\,$\pm$\,0.23\\
 187353    & 7300\,$\pm$\,230 & 7200\,$\pm$\,100 & 4.1\,$\pm$\,0.1 & 1.7\,$\pm$\,0.1 & 35\,$\pm$\,2  & 7.36\,$\pm$\,0.22\\
 206043    & 7200\,$\pm$\,190 & 7200\,$\pm$\,200 & 4.0\,$\pm$\,0.2 & 2.5\,$\pm$\,0.2 & 135\,$\pm$\,5 & 7.50\,$\pm$\,0.23\\
 075202    & 7700\,$\pm$\,260 & 7900\,$\pm$\,200 & 4.2\,$\pm$\,0.2 & 0.4\,$\pm$\,0.2 & 104\,$\pm$\,2 & 7.51\,$\pm$\,0.26\\
 091201    & 7100\,$\pm$\,150 & 7100\,$\pm$\,100 & 3.8\,$\pm$\,0.2 & 2.3\,$\pm$\,0.1 & 50\,$\pm$\,1  & 7.50\,$\pm$\,0.18\\
 103257    & 6900\,$\pm$\,160  & 7100\,$\pm$\,200 & 4.0\,$\pm$\,0.2 & 2.3\,$\pm$\,0.2 & 70\,$\pm$\,2  & 7.31\,$\pm$\,0.20\\
 113357    & 7000\,$\pm$\,160  & 7100\,$\pm$\,100 & 4.1\,$\pm$\,0.1 & 2.9\,$\pm$\,0.2 & 67\,$\pm$\,1  & 7.28\,$\pm$\,0.19\\
 133803    & 7000\,$\pm$\,170  & 7000\,$\pm$\,100 & 4.2\,$\pm$\,0.3 & 2.2\,$\pm$\,0.2 & 92\,$\pm$\,2  & 7.37\,$\pm$\,0.18\\
 137785    & 7000\,$\pm$\,170 & 6900\,$\pm$\,100 & 3.8\,$\pm$\,0.2 & 2.8\,$\pm$\,0.2 & 109\,$\pm$\,3 & 7.16\,$\pm$\,0.18\\
 149989    & 7000\,$\pm$\,190 & 7100\,$\pm$\,100 & 4.0\,$\pm$\,0.2 & 2.8\,$\pm$\,0.2 & 140\,$\pm$\,6 & 7.09\,$\pm$\,0.19 \\
 188032    & 7000\,$\pm$\,160  & 7100\,$\pm$\,100 & 4.0\,$\pm$\,0.1 & 2.5\,$\pm$\,0.2 & 54\,$\pm$\,2  & 7.34\,$\pm$\,0.18\\
 197451    & 7400\,$\pm$\,230 & 7300\,$\pm$\,200 & 4.0\,$\pm$\,0.1 & 3.1\,$\pm$\,0.2 & 26\,$\pm$\,3  & 7.73\,$\pm$\,0.22\\
 206481    & 6900\,$\pm$\,170  & 6900\,$\pm$\,100 & 4.1\,$\pm$\,0.2 & 1.5\,$\pm$\,0.2 & 86\,$\pm$\,2  & 7.36\,$\pm$\,0.18\\
 224288    & 7100\,$\pm$\,150  & 7100\,$\pm$\,200 & 3.9\,$\pm$\,0.1 & 2.2\,$\pm$\,0.2 & 48\,$\pm$\,2  & 7.39\,$\pm$\,0.19\\
 112934    & 7000\,$\pm$\,170  & 7100\,$\pm$\,100 & 3.9\,$\pm$\,0.2 & 2.4\,$\pm$\,0.2 & 75\,$\pm$\,2  & 7.03\,$\pm$\,0.22\\
 115466    & 6800\,$\pm$\,150 & 7100\,$\pm$\,100 & 4.0\,$\pm$\,0.2 & 2.0\,$\pm$\,0.2 & 40\,$\pm$\,3  & 7.56\,$\pm$\,0.20\\
 124248    & 7000\,$\pm$\,150 & 7100\,$\pm$\,100 & 4.1\,$\pm$\,0.1 & 1.7\,$\pm$\,0.2 & 50\,$\pm$\,3  & 7.37\,$\pm$\,0.20\\
 171834   & 6700\,$\pm$\,170 & 7000\,$\pm$\,100 & 4.0\,$\pm$\,0.2 & 2.7\,$\pm$\,0.2 & 72\,$\pm$\,2  & 7.40\,$\pm$\,0.21\\
 172416    & 6400\,$\pm$\,150  & 6400\,$\pm$\,100 & 3.9\,$\pm$\,0.1 & 1.9\,$\pm$\,0.2 & 55\,$\pm$\,3  & 7.41\,$\pm$\,0.20\\
 175337    & 6900\,$\pm$\,150 & 7100\,$\pm$\,100 & 4.0\,$\pm$\,0.1 & 1.7\,$\pm$\,0.1 & 38\,$\pm$\,2  & 7.73\,$\pm$\,0.19\\
 187028    & 6900\,$\pm$\,170 & 7300\,$\pm$\,200 & 4.5\,$\pm$\,0.2 & 2.3\,$\pm$\,0.3 & 87\,$\pm$\,3  & 7.23\,$\pm$\,0.23\\
 209295*   & 7400\,$\pm$\,170 & 7300\,$\pm$\,100 & 4.2\,$\pm$\,0.1 & 2.3\,$\pm$\,0.2 & 89\,$\pm$\,5  & 7.07\,$\pm$\,0.21 \\
 216910    & 6900\,$\pm$\,180  & 7100\,$\pm$\,100 & 4.3\,$\pm$\,0.2 & 2.1\,$\pm$\,0.2 & 95\,$\pm$\,4  & 7.66\,$\pm$\,0.21 \\
 224638    & 6900\,$\pm$\,140 & 7000\,$\pm$\,100  & 4.0\,$\pm$\,0.1& 1.5\,$\pm$\,0.2  & 29\,$\pm$\,7 & 7.39\,$\pm$\,0.20 \\
 224945    & 7000\,$\pm$\,150 & 7300\,$\pm$\,100 & 4.2\,$\pm$\,0.1 & 2.3\,$\pm$\,0.2 & 58\,$\pm$\,2  & 7.39\,$\pm$\,0.23\\
 041448    & 7300\,$\pm$\,170 & 7200\,$\pm$\,100 & 4.1\,$\pm$\,0.2 & 2.8\,$\pm$\,0.2 & 104\,$\pm$\,3 & 7.35\,$\pm$\,0.18 \\
  \hline
\end{tabular}

\begin{description}
 \item[*] Previously determined spectroscopic atmospheric parameters:
 \item[ ] \textit{\textbf{HD23585}}: \teff\,=7440 K,  \logg\,=4.29, $\xi$=3.0\,km\,s$^{-1}$ \citep{2001AJ....121.2159G}
 \item[ ] \textit{\textbf{HD26298}}: \teff\,=6790\,$\pm$\,200 K, \logg\,=3.95\,$\pm$\,0.22, $\xi$=1.5\,$\pm$\,0.5\,km\,s$^{-1}$, \textit{\textbf{HD110379}}: \teff\,=7140\,$\pm$\,160 K,  \logg\,=4.21\,$\pm$\,0.02, 
 \item[ ] $\xi$=1.5\,$\pm$\,0.4\,km\,s$^{-1}$, \textit{\textbf{HD126516}}: \teff\,=6590\,$\pm$\,120 K, \logg\,=4.01\,$\pm$\,0.15, $\xi$=1.9\,$\pm$\,0.3\,km\,s$^{-1}$ \citep{2008A&A...478..487B}
 \item[ ] \textit{\textbf{HD33204}}: \teff\,=7646 K,  \logg\,=4.11, $\xi$=3.4, \citep{1999A&A...351..247V}
 \item[ ] \textit{\textbf{HD46304}}: \teff\,=7048\,$\pm$\,16 K, \textit{\textbf{HD63436}}: \teff\,=6970 K, \logg\,=4.14, \textit{\textbf{HD106103}}: \teff\,=6610 K 
 \item[ ] \citep{2013A&A...553A..95M}
 \item[ ] \textit{\textbf{HD99329}}: \teff\,=6990\,K, \textit{\textbf{HD112934}}: \teff\,=7035 K, \textit{\textbf{HD209295}}: \teff\,=7392 K \citep{2012A&A...542A.116A}
 \item[ ] \textit{\textbf{H109799}}: \teff\,=6926\,$\pm$\,26 K \citep{2005PASP..117..911K}
 \end{description}

\end{table*} 

\begin{table*}
\centering
  \caption{The average abundances and standard deviations of individual elements of the stars. Number of the analysed parts is given in the brackets. The full table is available in the electronic form. }
  \begin{tabular}{llllll}
  \hline
  Elements & HD\,9565         & HD\,19655             & HD\,21788             & HD\,22702            &HD\,23005		  \\ 
(Atomic number)               &                       &                       &                      & 			    \\
 \hline 
C  (6)  &8.48\,$\pm$\,0.15 (7) & 8.34\,$\pm$\,0.24 (2)&8.52\,$\pm$\,0.21 (19) &8.65\,$\pm$\,0.01 (3) &8.11\,$\pm$\,0.27 (8) \\
N (7)  &                      &		              &8.61\,$\pm$\,0.19 (1)  &		             &			   \\
O (8)  &                      &   		      &8.57\,$\pm$\,0.19 (2)  &		             &			   \\
Na (11)&5.62\,$\pm$\,0.14 (2) &6.65\,$\pm$\,0.24 (1)  &6.08\,$\pm$\,0.05 (4)  &6.25\,$\pm$\,0.13 (1) &6.47\,$\pm$\,0.28(2)  \\
Mg (12)&7.64\,$\pm$\,0.09 (8) &7.57\,$\pm$\,0.01 (4)  &7.51\,$\pm$\,0.22 (11) &7.69\,$\pm$\,0.15 (5) &7.86\,$\pm$\,0.20 (8) \\
Si (14)&7.21\,$\pm$\,0.30 (16)&6.48\,$\pm$\,0.35 (4)  &7.09\,$\pm$\,0.43 (40) &6.94\,$\pm$\,0.32 (6) &7.15\,$\pm$\,0.35 (21)\\
S (16) &7.29\,$\pm$\,0.14 (2) &		              &7.20\,$\pm$\,0.12 (8)  &		             &7.41\,$\pm$\,0.28 (2)\\
Ca (20)&6.48\,$\pm$\,0.19 (19)&6.32\,$\pm$\,0.16 (4)  &6.38\,$\pm$\,0.20 (32) &6.29\,$\pm$\,0.06 (4) &7.01\,$\pm$\,0.30 (27)\\
Sc (21)&2.94\,$\pm$\,0.12 (10)&3.44\,$\pm$\,0.24 (5)  &3.20\,$\pm$\,0.07 (8)  &3.52\,$\pm$\,0.13 (2) &3.47\,$\pm$\,0.30 (12)\\
Ti (22)&4.94\,$\pm$\,0.11 (23)&4.94\,$\pm$\,0.04 (7)  &4.95\,$\pm$\,0.23 (92) &4.85\,$\pm$\,0.16 (9) &4.93\,$\pm$\,0.23 (36)\\
V (23) &4.94\,$\pm$\,0.14 (2) &		              &3.88\,$\pm$\,0.21 (10) &		             &4.26\,$\pm$\,0.50 (4) \\
Cr (24)&5.56\,$\pm$\,0.07 (15)&5.59\,$\pm$\,0.15 (7)  &5.48\,$\pm$\,0.23 (85) &5.57\,$\pm$\,0.06 (3) &5.64\,$\pm$\,0.44 (39)\\
Mn (25)&5.03\,$\pm$\,0.33 (8) &4.85\,$\pm$\,0.24 (1)  &4.94\,$\pm$\,0.25 (21) &5.17\,$\pm$\,0.13 (2) &5.30\,$\pm$\,0.19 (8) \\
Fe (26)&7.39\,$\pm$\,0.06 (40)&7.32\,$\pm$\,0.04 (9)  &7.26\,$\pm$\,0.14 (299)&7.40\,$\pm$\,0.13 (15)&7.61\,$\pm$\,0.12 (127\\
Co (27)&                      &		              &4.40\,$\pm$\,0.21 (6)  &		             &5.37\,$\pm$\,0.28 (4) \\
Ni (28)&6.13\,$\pm$\,0.08 (16)&6.03\,$\pm$\,0.08 (3)  &5.94\,$\pm$\,0.16 (88) &6.20\,$\pm$\,0.13 (2) &6.41\,$\pm$\,0.20 (23)\\
Cu (29)&3.56\,$\pm$\,0.14 (2) &		              &3.52\,$\pm$\,0.24 (3)  &		             &4.11\,$\pm$\,0.28 (1) \\
Zn (30)&                      &		              &4.18\,$\pm$\,0.19 (1)  &		             &			   \\
Sr (38)&3.06\,$\pm$\,0.14 (1) &		              &3.73\,$\pm$\,0.19 (1)  &2.39\,$\pm$\,0.13 (1) &3.92\,$\pm$\,0.28 (2) \\
Y (39) &2.31\,$\pm$\,0.13 (4) &1.92\,$\pm$\,0.24 (2)  &2.34\,$\pm$\,0.10 (10) &2.61\,$\pm$\,0.13 (2) &3.17\,$\pm$\,0.23 (6) \\
Zr (40)&2.43\,$\pm$\,0.14 (2) &2.62\,$\pm$\,0.24 (1)  &2.92\,$\pm$\,0.21 (12) &2.80\,$\pm$\,0.13 (2) &3.16\,$\pm$\,0.28 (2) \\
Ba (56)&2.27\,$\pm$\,0.16 (3) &2.98\,$\pm$\,0.24 (2)  &2.95\,$\pm$\,0.15 (4)  &2.57\,$\pm$\,0.13 (2) &2.77\,$\pm$\,0.34 (3) \\
La (57)&                      & 		      &1.56\,$\pm$\,0.19 (2)  &		             &2.11\,$\pm$\,0.28 (2) \\
Ce (58)&	              &		              &1.79\,$\pm$\,0.09 (10) &		             &1.84\,$\pm$\,0.28 (2) \\
Pr (59)&	              &		              &0.33\,$\pm$\,0.19 (1)  &		             &			   \\
Nd (60)&	              &		              &1.54\,$\pm$\,0.20 (19) &		             &1.59\,$\pm$\,0.28 (2) \\
Sm (62)&	              &		              &1.54\,$\pm$\,0.19 (1)  &		             &			   \\
 
\hline
\end{tabular}
\end{table*}

\end{document}